\documentclass[%
 reprint,
superscriptaddress,
nofootinbib,
 amsmath,amssymb,
 aps,
floatfix,
]{revtex4-2}
\usepackage{hyperref}
\usepackage{amssymb}
\usepackage{epsfig,amsmath,graphics}
\usepackage{pstricks}
\usepackage{color}
\usepackage{dcolumn}
\usepackage{xcolor}
\usepackage{tikz-feynman}[luatex=true]
\usepackage{comment}
\usepackage{slashed}
\usepackage[nameinlink]{cleveref}
\Crefname{figure}{Fig.}{Figs.}
\usepackage{graphicx}
\usepackage{float}
\usepackage{enumitem}
\usepackage{tabularx}

\definecolor{linkcolor}{rgb}{0.0, 0.28, 0.67}

\newcommand{\Mpc}{\text{Mpc}}

\def\nir{$N_{\rm IR}\,$}

\newcommand{\be}{\begin{equation}}
\newcommand{\ee}{\end{equation}}

\def\bea{\begin{eqnarray}}
\def\eea{\end{eqnarray}}
\def\ltap{\ \raise.3ex\hbox{$<$\kern-.75em\lower1ex\hbox{$\sim$}}\ }
\def\gtap{\ \raise.3ex\hbox{$>$\kern-.75em\lower1ex\hbox{$\sim$}}\ }
\def\lsim{\ \raise.3ex\hbox{$<$\kern-.75em\lower1ex\hbox{$\sim$}}\ }
\def\gsim{\ \raise.3ex\hbox{$>$\kern-.75em\lower1ex\hbox{$\sim$}}\ }

\newcommand{\ignore}[1]{}
\newcommand{\beq}{\begin{equation}}
\newcommand{\eeq}{\end{equation}}
\newcommand{\bear}{\begin{eqnarray}}
\newcommand{\eear}{\end{eqnarray}}

\def\lcdm{$\Lambda{\rm CDM}\,$}

\newcommand{\Ho}{$H_0\,$}
\newcommand{\Neff}{$N_{\rm eff}\,$}
\newcommand{\dataD}{$\mathcal{D}\,$}
\newcommand{\dataDH}{$\mathcal{D}H\,$}
\newcommand{\dataDL}{$\mathcal{D}L\,$}
\newcommand{\dataDHL}{$\mathcal{D}HL\,$}
\newcommand{\lya}{Ly$\alpha\,$}

\begin{document}

\title{Stepping into the Forest: Confronting Interacting Radiation Models for the Hubble Tension with Lyman-$\alpha$ Data}

\author{Hengameh Bagherian}
\affiliation{Department of Physics, Harvard University, Cambridge, MA 02138, USA}
\author{Melissa Joseph}
\affiliation{Department of Physics and Astronomy, University of Utah, Salt Lake City, UT 84112, USA}
\author{Martin Schmaltz}
\affiliation{Physics Department, Boston University, Boston, MA 02215, USA}
\affiliation{Center for Cosmology and Particle Physics, Department of Physics, New York University, New York, NY 10003, USA}
\affiliation{Institute for Theoretical Physics, Georg-August University Göttingen, 37077 Göttingen, Germany}
\author{Eashwar N. Sivarajan}
\affiliation{Physics Department, Boston University, Boston, MA 02215, USA}

\begin{abstract}
    Models of interacting dark radiation have been shown to alleviate the Hubble tension. Extensions incorporating a coupling between dark matter and dark radiation (DM-DR) have been proposed as combined solutions to both the Hubble and $S_8$ tensions. A key feature of these extended models is a break in the matter power spectrum (MPS), suppressing power for modes that enter the horizon before the DM-DR interactions turn off. In scenarios with a massless mediator, modes that enter before matter-radiation equality get suppressed, whereas for massive mediators, the break is determined by the mediator mass, a free parameter. In this work, we test these models against probes of LSS: weak lensing, CMB lensing, full-shape galaxy clustering, and eBOSS measurements of the 1D \lya forest flux power spectrum. The latter are the most constraining since they probe small scales where many models predict the largest deviations. In fact, already within \lcdm, the eBOSS \lya data are in significant tension with Planck CMB data, with the \lya data preferring a steeper slope of the MPS at $k \sim h \mathrm{Mpc}^{-1}$. We find that the simplest dark radiation models, which improve the Hubble tension, worsen the fit to the \lya data. However, models with DM-DR interactions can simultaneously address both tensions.
    \end{abstract}

\maketitle

\section{Introduction}
\label{sec:introduction}

The ``Hubble tension,'' a discordance between different determinations of the expansion rate of the universe today (\Ho), is perhaps the most significant challenge to the current cosmological model (\lcdm). Direct local measurements of \Ho show approximately 5-10\% more rapid expansion than the values measured from the CMB and from BAO, which both rely on the \lcdm model. A very simple and sharp formulation of the Hubble tension, which we will adopt for this paper, is the $\sim 5 \sigma$ discrepancy between the local SH0ES~\cite{Riess:2021jrx} determination of \Ho using calibrated supernovae and the \lcdm determination of \Ho from a global fit to Planck CMB data~\cite{Planck:2018vyg}.
However, we emphasize that the tension is robust to replacing the local SH0ES measurement with another local measurement (in particular TRGB~\cite{Freedman:2019jwv,Scolnic:2023mrv}) or replacing the Planck CMB data in the \lcdm fit with CMB measurements by ACT~\cite{ACT:2020gnv} or SPT~\cite{SPT-3G:2021eoc,SPT-3G:2022hvq} or using BAO~\cite{Schoneberg:2019wmt, DESI:2024mwx} instead.

Of the many solutions to the Hubble tension that have been proposed, those that seek to modify the sound horizon at recombination, $r_s$, by modifying \lcdm at early times before recombination appear particularly interesting~\cite{Aylor:2018drw}. Such solutions avoid late modifications of the expansion history which are severely constrained by PANTHEON data~\cite{Pantheon}. They also explain why CMB and BAO based measurements which both rely on $r_s$ agree with each other but disagree with direct measurements of \Ho.
However, any modifications to \lcdm before recombination are severely constrained by the high precision of CMB data. In fact, the \Ho values from CMB and supernova measurements are more than 10$\sigma$ apart when measured in units of the 1$\sigma$ Planck2018 errors. This makes solutions to the Hubble tension which modify the CMB fit with new early universe physics challenging.

The angular sound horizon at recombination, $\theta_s$, and the redshift of equality, $z_{\rm eq}$, are especially well determined by the CMB data, and any new physics must not shift them significantly from their \lcdm values. 
The angular sound horizon is determined from the comoving sound horizon $r_s$ as $\theta_s = r_s/D_A$, where $D_A = \int_{0}^{z_*}  \mathrm{d}z/H \propto H_0^{-1}$. Thus a decrease in $r_s$ due to additional energy density beyond \lcdm before recombination requires a compensating decrease in $D_A$ which implies a larger value for \Ho. This is how additional early energy density increases \Ho in CMB fits and alleviates the Hubble tension.\footnote{An analogous argument applies to BAO with the redshift of recombination $z_*$ replaced by the redshift of the BAO measurement $z_{\rm BAO}$.} 

Two well-known classes of models for such additional early energy density are Early Dark Energy (EDE)~\cite{Poulin:2018cxd, Agrawal:2019lmo, Lin:2019qug, Poulin:2023lkg,Niedermann:2019olb} and dark radiation (\Neff)~\cite{Planck:2018vyg}. EDE models introduce additional energy density in the form of a finely-tuned new fluid whose energy density is time-independent until matter-radiation equality and then undergoes a transition to rapidly redshift away. 

In \Neff models, the additional energy density resides in dark radiation which does not require such a finely-tuned transition because radiation energy density automatically redshifts away after matter-radiation equality. A number of different variants of \Neff models have been proposed in which the dark radiation may be free-streaming or self-interacting~\cite{Blinov:2020hmc, Brust:2017nmv,Escudero:2021rfi}. Interacting dark radiation sectors can also contain multiple particle species with different small masses. As the temperature drops below a mass threshold in such a dark sector, its energy density increases, leading to a \textit{step} in \Neff~\cite{Aloni:2021eaq}. 

Both EDE and \Neff models alleviate the Hubble tension while maintaining good fits to CMB and BAO data. However, it was found that models with sufficient EDE to significantly improve the Hubble tension are in conflict with large scale structure (LSS) data~\cite{Hill:2020osr,Ivanov:2020ril}.  In particular, the full-shape MPS measurement from galaxy clustering data from the Baryon Oscillation Spectroscopic Survey (BOSS)~\cite{Ivanov:2019pdj, Zhang:2021yna} and measurements of the amplitude of density fluctuations on scales $\sim 8 h^{-1}$ Mpc (or $\sigma_8$) extracted from weak lensing surveys~\cite{DES:2021wwk,Heymans:2020gsg} constrain the allowed amount of EDE when included in global fits. In addition,~\cite{Goldstein:2023gnw} found that the EDE solution to the Hubble tension significantly worsens the already poor \lcdm fit to \lya data. On the other hand,~\cite{Cruz:2023lmn,Niedermann:2020qbw} showed that the LSS tensions can potentially be resolved with additional fields and dynamics beyond EDE.

In this paper, we investigate how several recently proposed \Neff solutions to the Hubble tension hold up when confronted with full-shape LSS, $S_8$, and \lya data. Among this data, \lya is unique as it probes the smallest scales, which are sensitive to the earliest new physics affecting the growth of structure and the MPS. The \lya forest is formed when photons from distant quasars are absorbed by neutral hydrogen in the intergalactic medium at redshifts $2 < z < 5$, offering insights into matter clustering at higher redshifts and smaller scales than weak lensing and galaxy clustering data. The results from such surveys~\cite{eBOSS:2018qyj, Irsic:2017sop} are measurements of the non-linear MPS for a range of wavenumbers $k$ in bins of redshift, which are hard to interpret without simulations of non-linear matter clustering and knowledge of bias parameters. We will use a more \textit{user-friendly} summary of the data in the form of a compressed likelihood for the amplitude and slope of the \textit{linear} MPS at $k = 0.009 \text{ (km/s)}^{-1}$ and $z= 3$. Interestingly, the two measurements of the \lya forest, eBOSS and XQ-100, show $3-5 \sigma$ tensions with \lcdm fit to the CMB, with both preferring a significantly steeper slope of the MPS at $\sim$ Mpc scales than the best fit \lcdm model~\cite{Rogers:2023upm}.

\begin{figure}[th!]
	\includegraphics[width=0.48\textwidth]{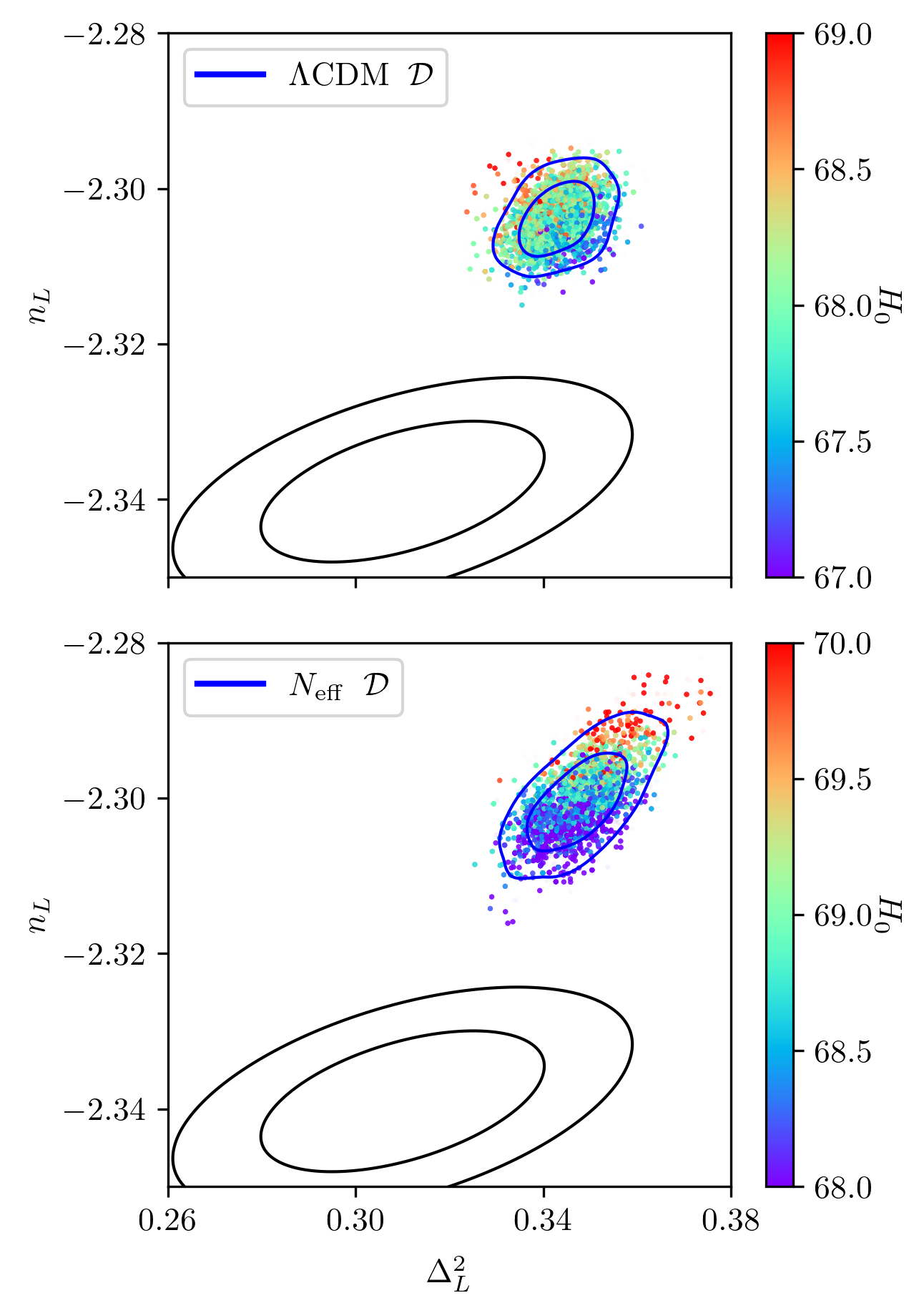}
	\caption{\textbf{(Top)} Marginalized posterior distributions for \lya likelihood parameters, $\Delta_L^2$ and $n_L$, from \lcdm model fits to the \dataD dataset (blue contours, dataset \dataD is defined in Section~\ref{sec: methods}) are shown with \Ho indicated by a color gradient. Solid black contours represent the $68\%$ and $95\%$ confidence regions from the eBOSS \lya likelihood. \textbf{(Bottom)} Similarly, for the \lcdm +\Neff model, highlighting the difficulty of solving the Hubble tension while remaining consistent with the eBOSS \lya data.}
	\label{fig:1}
\end{figure}

An analysis of EDE using \lya data finds that the parameter space in EDE models which alleviates the Hubble tension \textit{increases} the tension with the \lya data~\cite{Goldstein:2023gnw}. The potential for similar trouble in extra radiation models is straightforward to understand. In the simplest model of $\Lambda$CDM+$N_{\rm eff}$ the additional free-streaming radiation (at fixed $\theta_s$ and $z_{eq}$) increases the diffusion damping of the high-$l$ tail of the CMB. To partially compensate this effect a global fit will increase the primordial tilt $n_s$. However, this also increases the slope of the matter perturbations, making the disagreement with LSS and especially the \lya data worse. The resulting tension is evident in Fig.~\ref{fig:1}, which compares the preferred regions of \lya data from eBOSS~\cite{eBOSS:2018qyj} in the slope $n_L$ versus amplitude $\Delta_L^2$ plane with the posteriors of \lcdm (top panel) and \lcdm+$N_{\rm eff}$ (bottom panel) derived from fits to CMB+BAO+Pantheon (uncalibrated supernovae) data.

When observing significant discrepancies between two separate measurements, caution is warranted in drawing conclusions from a combined analysis. One plausible explanation is that one of the measurements may have a large hidden systematic error. Alternatively, the discrepancy could result from large statistical fluctuations. This type of error typically diminishes with extended data collection as per the law of large numbers. For the specific case of a $\sim 5\sigma$ tension between Planck CMB and \lya in \lcdm, a more probable explanation than statistical fluctuation might be the use of an incorrect cosmological model to fit the data. However, note that for both \lcdm and \lcdm+$N_{\rm eff}$ the tension with the \lya data increases with increasing \Ho.

For the remainder of this paper, we assume the datasets are accurate, unaffected by systematic errors, and we investigate the consistency of models of interacting dark sectors which significantly lessen the Hubble tension with the LSS and \lya data. Our main results include:
\begin{itemize}
	\item The simplest models of interacting radiation, with (WZDR) or without (SIDR) a step, which yield larger values of \Ho, align with the $S_8$ values from weak lensing and full-shape LSS data. However, they exacerbate the existing tension with \lya data (see Fig.~\ref{fig:3}). Future \lya data, if confirming the current preference for a suppressed slope of the MPS, could rule these models out.
	\item Models where the interacting dark radiation also couples to dark matter (SIDR+, WZDR+) show consistency with all LSS data, including \lya (see Fig.~\ref{fig:5}), while still significantly alleviating the Hubble tension.
\end{itemize}


\section{Interacting dark radiation models}
\label{sec:models}


Models with interacting dark radiation are less constrained by CMB fits (\cite{Blinov:2020hmc}) than models with similar amounts of free-streaming radiation because interacting radiation is less efficient at erasing small-scale structure.
Therefore, models with interacting radiation can reduce the Hubble tension better than those with free-streaming radiation (see~\cite{Aloni:2021eaq,Brinckmann:2022ajr,Allali:2024anb,Allali:2024cji,Ghosh:2021axu,Ghosh:2024wva} for detailed studies).

A significant additional constraint on the allowed amount of extra dark radiation is provided by BBN. However, for interacting dark radiation, it is reasonable to assume that the dark sector is populated through thermalization with SM neutrinos~\cite{Aloni:2023tff,Berlin:2017ftj,Berlin:2018ztp,Berlin:2019pbq}. The constraints from BBN can be completely bypassed if this thermalization occurs below $\sim$ 100 keV~\cite{Berlin:2017ftj, Aloni:2023tff, Giovanetti:2024orj,Garny:2024ums}.\footnote{Combined CMB and BBN constraints on extra radiation present at BBN are particularly stringent because an effective number of neutrino species \Neff$>3.043$ during BBN leads to an increased helium abundance. This increased helium abundance worsens the fit to the CMB for models with extra radiation by enhancing Silk damping, in addition to the damping of the CMB tail already caused by the extra radiation.} Therefore, this work focuses only on models in which \Neff is similar to that of \lcdm at BBN and increases between BBN and the CMB. The mechanism behind this increase, whether from scattering and mixing with neutrinos or from a late decay, is not central to our analysis; we simply assume it occurs and establishes the early-time value of \Neff. In~\ref{subsec:models_summary}, we provide a brief summary of the models analyzed in this paper, while~\ref{subsec:models_detail} offers more detailed information on each model. Readers already familiar with these models may choose to skip this section. 

\subsection{Summary of models considered}
\label{subsec:models_summary}

\begin{figure}[t!]
    \centering
    \includegraphics[width=0.49\textwidth]{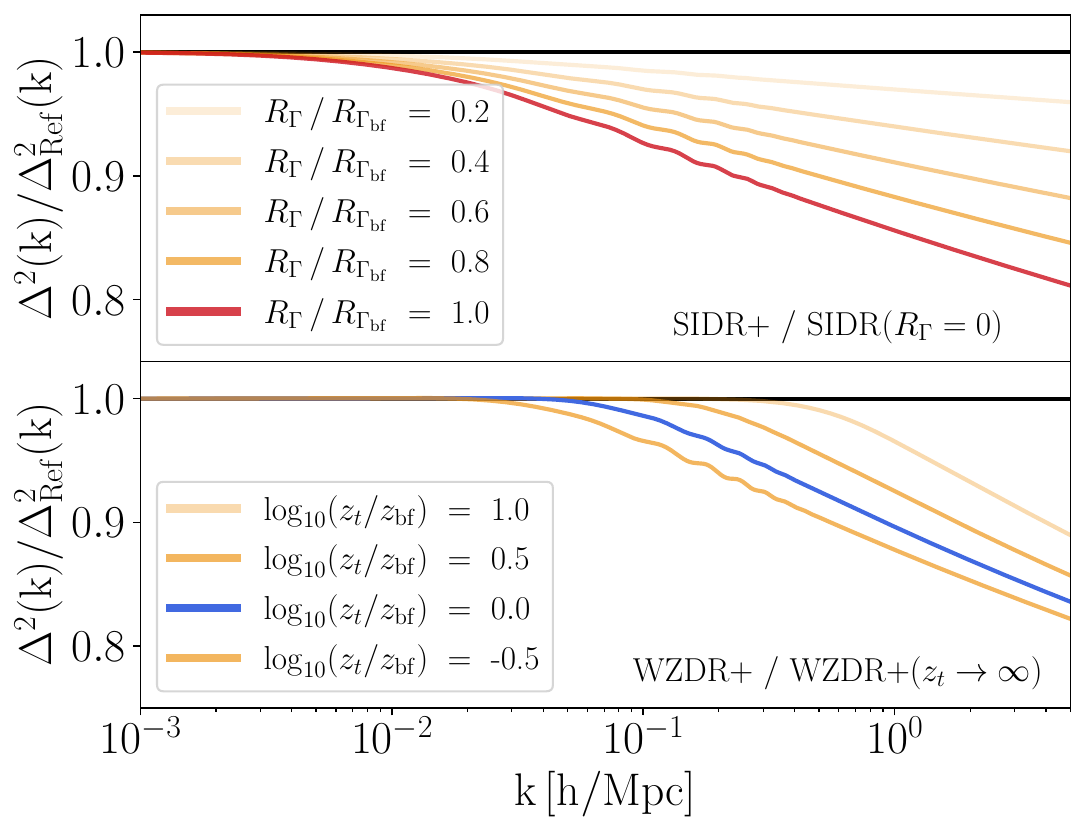}
    \caption{\textbf{(Top)} This ratio plot compares the MPS of the SIDR+ model to a non-interacting SIDR model, highlighting how increased interaction strength $R_{\Gamma}$ enhances MPS damping at scales smaller than $k_{eq} \sim 10{-2}$ h/Mpc. The red line indicates the best-fit parameters of the SIDR+ model to the \dataDHL dataset. \textbf{(Bottom)} This plot contrasts the MPS of the WZDR+ model with an early transition redshift model, illustrating how $z_t$ shifts the suppression scale horizontally in the WZDR+ model. The blue line represents the WZDR+ model's best fit to the \dataDHL dataset.}\label{fig:2} 
\end{figure}
\begin{itemize}
\item
\textit{Strongly Interacting Dark Radiation (SIDR)}: This model posits a single self-interacting massless particle that becomes thermalized after BBN. 
It provides an excellent fit to the CMB data while also accommodating higher values of \Ho\cite{Blinov:2020hmc}. SIDR introduces one additional cosmological parameter compared to \lcdm: \nir, which quantifies the amount of extra dark radiation.

\item
\textit{Wess-Zumino Dark Radiation (WZDR)}\cite{Aloni:2021eaq}:
WZDR extends SIDR by adding a small mass to one of the dark radiation particles. As the dark temperature drops below this mass threshold, the massive particles annihilate away, thereby increasing the total energy density in the dark radiation relative to a model without such a \textit{step} in \Neff. This model is characterized by \nir and $z_t$, the former corresponds to the abundance of interacting radiation after the step, while the latter indicates the transition redshift at which the step occurs, determined by the particle mass.

\item 
\textit{SIDR with dark matter - dark radiation interactions (SIDR+)}\cite{Buen-Abad:2015ova,Lesgourgues:2015wza,Pan:2018zha}:
Motivated by our result that neither SIDR nor WZDR significantly alleviates the tension between the CMB-derived MPS slope and the \lya data (see Fig.~\ref{fig:3}), we explore models with dark matter -- dark radiation interactions (DM-DR). One such model, named SIDR+ for its addition of DM-DR interactions to the SIDR framework, features self-interacting massless scalars and a Yukawa coupling to dark matter fermions. SIDR+ introduces another cosmological parameter, the momentum transfer ratio $R_{\Gamma}$, which is a measure of the strength of the DM-DR interaction. The top panel of Fig.~\ref{fig:2} shows that increasing $R_{\Gamma}$ smoothly suppresses the MPS at high $k>k_{\rm eq}\,$.
\item 
\textit{WZDR with dark matter - dark radiation interactions (WZDR+)}~\cite{Joseph:2022jsf,Schoneberg:2023rnx}:
Adding a coupling between DM-DR in the WZDR model also attenuates the MPS tail; we call this extension WZDR+. The implementation of WZDR+ is similar to that of SIDR+ with the difference that the dark matter is now only directly coupled to the massive component of the interacting fluid. Below redshift $z_t$, the DM-DR momentum transfer rate becomes suppressed by the mediator mass, effectively ending the DM-DR interaction. Thus WZDR+ has three parameters beyond those of \lcdm: \nir, $R_{\Gamma}$, and $z_t$. The latter determines the onset of MPS suppression, as shown in the bottom panel of Fig.~\ref{fig:2}.
\end{itemize}

\subsection{Details of models considered}\label{subsec:models_detail}

In this Section, we delve into the specifics of how each model addresses the Hubble and \lya tensions and the rationale for introducing each additional cosmological parameter. We begin with the SIDR model:

\begin{itemize}

\item
\textit{SIDR}: 
SIDR improves on several key issues associated with free-streaming extra radiation. Free-streaming radiation hinders structure formation due to enhanced diffusion damping and shifts the CMB peaks to higher multipoles ($\ell$ values) through the \textit{neutrino drag effect}~\cite{Bashinsky:2003tk}. Furthermore, CMB modes entering the horizon during radiation domination experience an increase in amplitude, known as the \textit{radiation driving effect}, which is inversely proportional to $1+\frac{4}{15}\rho_{\rm fs}/\rho_{\rm tot}$~\cite{Bashinsky:2003tk, Ma:1995ey, Blinov:2020hmc}. 

The presence of free-streaming radiation diminishes this amplitude boost for $\ell \geq \ell_\mathrm{eq}$ by increasing the fraction of free-streaming species. Consequently, the Planck CMB data places strong constraints on the amount of free-streaming radiation.

SIDR, a perfect fluid with an equation of state $w=1/3$ and sound speed $1/\sqrt{3}$, overcomes limitations by using perfect fluid dynamics. This approach addresses problems related to free-streaming radiation, enabling more radiation for a better \Ho fit to SH0ES while still fitting CMB data well.

\item \textit{WZDR}:
WZDR~\cite{Aloni:2021eaq} is a model featuring an interacting dark radiation fluid composed of two particle species in thermal equilibrium: one, $\phi$, is massive and annihilates into the other, $\psi$, which is massless, when the temperature in the sector drops below the mass of $\phi$. A simple realization of this scenario is provided by the Wess-Zumino model with soft supersymmetry breaking via the scalar mass $m_{\phi}$. The interactions in WZDR are:
\begin{equation}\label{eq:lagrangian}
  \mathcal{L}_{\rm{WZDR}} = \lambda \phi\bar{\psi}\psi+\lambda^2(\phi^*\phi)^2\,.
\end{equation}
The annihilating $\phi$ particles transfer their entropy to the massless $\psi$ particles, thereby increasing $N_{\rm eff}$ of the combined $\phi-\psi$ fluid. Assuming a sufficiently large $\lambda$, the remaining $\psi$ particles continue to scatter efficiently via a four-fermi interaction, ensuring the dark radiation remains a perfect fluid throughout cosmic expansion. The increase, or step, in $N_{\rm eff}$ is easy to understand because during the era of $\phi$ particle annihilation, at $z_t\,$, the dark radiation sector is actually a mix of radiation and matter. This mix's energy density redshifts more slowly than pure radiation. Additionally, the radiation and matter mix temporarily reduces the equation of state and sound speed of the combined fluid below 1/3.

The step in \Neff at $z_t$ also introduces a step in the background expansion rate, $H(z)$, leading to an $\ell$-dependent modulation of CMB peaks.  Modes that enter the horizon before the step experience a slower expansion rate, causing peaks at higher $\ell$ values to shift to slightly lower $\ell$, more effectively counteracting the drag effect than SIDR. Reference~\cite{Aloni:2021eaq} demonstrated that a step in \Neff near $z_t\sim 10^4$ enhances the fit to Planck data and predicts larger values of \Ho from the CMB fit.

\item \textit{SIDR+}:
The SIDR+ model differs from SIDR by the additional coupling between dark matter and dark radiation. This coupling allows momentum transfer between the two fluids, which produces an effective drag force on the DM fluid that slows the clustering of matter. Assuming that the momentum transfer rate scales with temperature as $\Gamma(T) \propto T^2$, as is the case in the SIDR+ model, the ratio $\Gamma/H$ is constant during radiation domination, which is key for smoothly suppressing the MPS for all momentum modes that enter the horizon before matter-radiation equality. During matter domination, the Hubble parameter's temperature dependence shifts to $H(T) \sim T^{3/2}$, effectively shutting off DM-DR interactions around equality. Thus, only modes that crossed the horizon before equality experience slower growth, producing a \textit{break} in the MPS at $\sim k_{\rm eq}\,$. Since the impact of the DM-DR interactions on cosmology takes place during radiation domination, when the ratio $R_{\Gamma} \equiv \Gamma/H$ is constant, we use $R_{\Gamma}$ during radiation domination as the additional parameter of SIDR+ . In prior literature~\cite{Buen-Abad:2017gxg,Pan:2018zha,Archidiacono:2019wdp,Joseph:2022jsf}, the interaction strength has also been parameterized with $\Gamma_0\,$, the strength of the interaction extrapolated to today via its $T^2$ scaling. The two definitions are related by $R_{\Gamma} = \Gamma_0 / (H_0 \sqrt{\Omega_r})\,$.

The break in the MPS, resulting from the suppressed growth of matter perturbations relative to \lcdm, is analytically approximated in Eq.~\ref{eq:slope}, with the slope of the decline equal to $-R_{\Gamma}$~\cite{Joseph:2022jsf}:
\begin{equation}\label{eq:slope}
    \Delta^2_{R_{\Gamma}\neq 0}\simeq \begin{cases}
        \Delta^2_{R_{\Gamma}= 0} \quad &k \ll k_{\rm eq}\\
        \Delta^2_{R_{\Gamma}= 0}\left(\frac{k}{k_{\rm eq}}\right)^{-R_{\Gamma}} \quad &k \gg k_{\rm eq}
    \end{cases}\,,
\end{equation}
where $k_{\rm eq} $ is the mode which enters the horizon during matter-radiation equality. 

As mentioned earlier, the increase in $H(z)$ due to the additional radiation in SIDR (and WZDR) also increase the amount of diffusion damping. In a global fit, this increased damping can be partially compensated for with a higher spectral tilt, $n_s$. However, this also boosts power in the MPS at smaller scales $\propto (k/k_{\rm piv})^{\Delta n_s}$. In the SIDR+ model this effect can be balanced by an adjustment in the interaction strength $R_{\Gamma}\sim\Delta n_s$ as per Eq.~\ref{eq:slope}.

This effect is somewhat similar to what happens in \lcdm extensions with a running spectral tilt, previously analyzed in the context of eBOSS \lya data by Rogers and Poulin~\cite{Rogers:2023upm}. Their analysis indicates that fitting the \lya data requires
\begin{equation}
\alpha_s \equiv \frac{dn_s}{d\ln k} = -0.0108 \pm 0.0022\,,
\end{equation}
which is much larger than expected in slow-roll inflaction for which one expects $\alpha_s = \mathcal{O}(|n_s-1|^2) \approx 0.001$.
Moreover, this degree of running for the tilt also modifies the MPS at scales sensitive to the CMB data. SIDR+, in contrast, introduces suppression only for $k>k_{\rm eq}$, leaving CMB observations largely unaffected. Furthermore, our approach does not face challenges from a model-building perspective since $R_{\Gamma}$ can natually take on values in a broad range. As we will show in subsequent sections, fits to LSS data prefer $R_{\Gamma}\sim \mathcal{O}(0.05)$.

The inclusion of $R_{\Gamma}$ as an additional parameter enables SIDR+ to better fit LSS data at small scales, accomodate a higher \Ho, and reduce the $S_8$ tension~\cite{Joseph:2022jsf} between the Planck CMB and LSS measurements. Fig.~\ref{fig:2} (Top) shows how SIDR+ more accurately fits the \dataDHL dataset than SIDR without DM-DR interaction, using $R_{\Gamma}$ to adjust the slope of the MPS at high $k$.

\item \textit{WZDR+}: WZDR+ adds a Yukawa interaction, $\lambda_2\, \phi\bar{\chi}\chi$, of the scalar $\phi$ to the DM particle $\chi$ to the WZDR Lagrangian~\ref{eq:lagrangian}. The t-channel $\chi\psi\rightarrow\chi\psi$ process through $\phi$ exchange dominates over all other DM-DR interactions, with its momentum transfer rate $\Gamma(T)\propto T^2$ when $\phi$ is relativistic, mirroring the SIDR+ scaling during radiation domination. Below temperatures of $m_{\phi}$, the momentum transfer rate from the four-fermi contact term $\bar{\psi}\psi\bar{\chi}\chi$ quickly becomes unimportant because it now scales as $\Gamma(T)\propto T^6$.

This means that the momentum exchange shuts off after crossing the $m_{\phi}$ threshold, or equivalently, below the transition redshift $1+z_t \equiv m_{\phi}/T_{d0}$. Consequently, the break in the MPS now occurs at $k_t$, with modes $k>k_t$ entering the horizon before the $m_{\phi}$ threshold being suppressed, while modes $k<k_t$ remain unaffected.

It should be noted that in a different version of WZDR+ with alternative couplings, the momentum transfer could occur between the DM and the massive radiation particles. In such cases, the exchange rate would decrease exponentially, $\propto e^{-m_{\phi}/T}$, as the massive particles annihilate away. This faster decoupling rate only slightly modifies cosmological fits. For more details on these models and alternative methods of suppressing the MPS, see~\cite{Buen-Abad:2022kgf,Buen-Abad:2023uva,Chacko:2016kgg,Archidiacono:2017slj,Chacko:2018vss,Garny:2018byk,Raveri:2017jto,Rubira:2022xhb,Archidiacono:2019wdp,Choi:2020pyy,Lehmann:2023vjv,Mazoun:2023kid,Zu:2023rmc}.

Fig.~\ref{fig:2} (Bottom) illustrates how $z_t$ affects the MPS in the WZDR+ model, showing the MPS for different $z_t$ values, normalized to a model where $z_t\rightarrow\infty$. Higher $z_t$ moves the transition to smaller scales but keeps larger scales, relevant to CMB data, completely unchanged. The interaction strength $R_{\Gamma}$ in WZDR+ is similar to SIDR+ (see top panel). The two additional degrees of freedom enable WZDR+ to align with LSS data and address the Hubble tension effectively.
\end{itemize}


\section{Data and MCMC}
\label{sec: methods}

We conduct an MCMC analysis on the models from~\ref{sec:models} using CLASS~\cite{Blas:2011rf} and MontePython~\cite{Brinckmann:2018cvx}, focusing on specific datasets. We use flat priors for the six \lcdm parameters $\{\omega_b,\, \omega_{\rm dm},\, \theta_s, n_s, A_s,\, \tau_{\rm reio} \}$\,. For SIDR+, we add flat priors for the dark radiation \nir $> 0.01$ the DM-DR interaction $R_{\Gamma}>0$. WZDR+ includes a logarithmic prior on the step's redshift $\log_{10} z_t \in [4.0, 4.6]$. As explained in the previous Section, we assume that the extra radiation in WZDR+ and SIDR+ occurs post-BBN, therefore leaving the BBN prediction for the primordial helium abundance $Y_p$ as in \lcdm (\Neff = 3.044). For generating figures and triangle plots, we used \texttt{GetDist}~\cite{Lewis:2019xzd}. 

Our baseline dataset $\boldsymbol{\mathcal{D}}$ comprises:
\begin{itemize}
	\item \textit{Planck 2018}~\cite{Planck:2018vyg}: Includes TT, TE, and EE data for both low-$\ell$ (`lowl\_TT', `lowl\_EE') and high-$\ell$ (`highl\_TTTEEE') ranges, along with the full set of nuisance parameters and the Planck lensing likelihood. 

    \item \textit{BAO+Pantheon}: Combines late-universe constraints with the BAO-only likelihood from BOSS DR12 ($z = 0.38, 0.51, 0.61$)~\cite{BOSS}, small-z BAO data from 6dF ($z = 0.106$)~\cite{6dF_fix} and MGS ($z = 0.15$)~\cite{MGS} catalogs, and the Pantheon supernova likelihood~\cite{Pantheon}. We did not include the more recent DES Supernova data~\cite{DES:2024tys} which prefer a high $\Omega_M= 0.352\pm0.017$.

    \item\textit{$S_8$}: This observable is measured through weak lensing and galaxy clustering, with the former including cosmic shear and galaxy-galaxy lensing. In this paper, we select the $S_8$ likelihood based on a combined cosmic shear analysis from the Dark Energy Survey (DES Y3) and the Kilo-Degree Survey (KiDS-1000), with $S_8 = 0.790^{+0.018}_{-0.014}$~\cite{Kilo-DegreeSurvey:2023gfr}, implemented as a Gaussian likelihood. We favor this measurement over one that includes galaxy clustering data because the EFTofBOSS likelihood which we include in dataset $\mathcal{D}$ relies on the BOSS DR12 galaxy sample analysis. Focusing solely on cosmic shear from weak gravitational lensing ensures the $S_8$ dataset's independence from galaxy clustering information in $\mathcal{D}$. Moreover, analyzing multiple surveys together provides a more precise $S_8$ value.

    \item \textit{EFTofBOSS}: Utilizes the Effective Field Theory of Large Scale Structure (EFTofLSS)~\cite{Zhang:2021yna} for analyzing the SDSS-III BOSS DR12 galaxy sample~\cite{BOSS,Kitaura:2015uqa}, incorporating both monopole and quadrupole moments of the galaxy power spectrum. Our analysis includes the Full Shape of the BOSS galaxy power spectrum and its cross-correlation with reconstructed BAO parameters~\cite{BOSS:2015fqm}, accessible via \texttt{Pybird}~\cite{DAmico:2020kxu}.\footnote{Class-PT is another code that computes the EFT of LSS for this likelihood, with minor differences in the priors for nuisance parameters. We find that the selection of priors on these nuisance parameters can result in significant overall changes to the $\chi^2$ values, while only causing negligible differences in the posteriors for the cosmological parameters.}
    \item\textit{EFTofeBOSS:} Analyzes eBOSS DR16 QSO data~\cite{eBOSS:2020yzd} with EFTofLSS predictions~\cite{Beutler:2021eqq,Simon:2022csv}. Covariance matrix derived from mock catalogs~\cite{eBOSS:2020mzp} using the Extended Zel'dovic approximation~\cite{Chuang:2014vfa}. Available via \texttt{Pybird}. 

    We refer to the combined datasets of EFTofBOSS and EFTofeBOSS as \textit{full-shape}.
\end{itemize} 

The dataset $\boldsymbol{\mathcal{H}}$ includes
\begin{itemize}
    \item \textit{SH0ES:} Measurement of the supernovae intrinsic magnitude $M_b=-19.253\pm0.027$ by Riess et al.~\cite{Riess:2021jrx}, included as a Gaussian likelihood.
\end{itemize}

The dataset $\boldsymbol{\mathcal{L}}$ originates from the one-dimensional Ly$\alpha$ forest flux power spectrum, utilizing data from the SDSS DR14 BOSS and eBOSS surveys~\cite{eBOSS:2018qyj}. The data spans thirteen redshift bins from $z \approx 2.2$ to $4.6$ and scales from $k \approx 0.001$ to $0.020 \text{ (km/s)}^{-1}$. For ease of comparison to models that modify the MPS, ~\cite{eBOSS:2018qyj} further provides in Fig. 20 the 2D posterior for the dimensionless amplitude of the linear power spectrum, $\Delta^2_L$, and its logarithmic slope, $n_L$, at the pivot wave number $k_p = 0.009\text{ (km/s)}^{-1}$ and redshift $z_p = 3$. The conversion from wavenumbers expressed in velocity units to comoving units is given by the factor $H(z_p)/(1+z_p)$ . We fit a 2D-Gaussian likelihood to the aforementioned posterior, which yielded:

\begin{itemize}
    \item $\Delta^2_L \equiv k^3 P_L/(2\pi^2) = 0.31 \pm 0.02$, $n_L \equiv \mathrm{d }\ln{P_L}/\mathrm{d }\ln{k} = -2.339 \pm 0.006$, with a correlation coefficient $r = 0.50$.  
\end{itemize}

Appendix~\ref{app:Lya_deets} supports the 2D-Gaussian likelihood's applicability to power-law extensions of \lcdm models, including SIDR+ and WZDR+.

A second Ly$\alpha$ likelihood was derived from observations of QSOs by XQ-100 at redshifts $z \approx 3.5-4.5$ and from MIKE/HIRES quasar samples~\cite{Irsic:2017sop,lopez2016xq,Esposito:2022plo,Viel:2013fqw}. This likelihood fits a 2D-Gaussian to the posteriors of $\Delta^2_L$ and $n_L$ from Fig.~A1 of Appendix A in~\cite{Esposito:2022plo}, as detailed in~\cite{Goldstein:2023gnw}. 
The XQ-100 likelihood differs from the eBOSS likelihood by covering a narrower redshift range and focusing on smaller, more non-linear scales. The results of the two analyses are in some tension with with each other. We anticipate the upcoming DESI 1D \lya flux power spectrum, expected within the next year, to provide clarity on the situation. In this paper, we fit only to the eBOSS \lya likelihood, though we also present XQ-100 contours for comparative purposes in Fig.~\ref{fig:5}.

\section{Results}\label{sec: results}
\subsection{The \lya analysis}

\begin{figure*}[!ht]
    \includegraphics[width=0.98\textwidth]{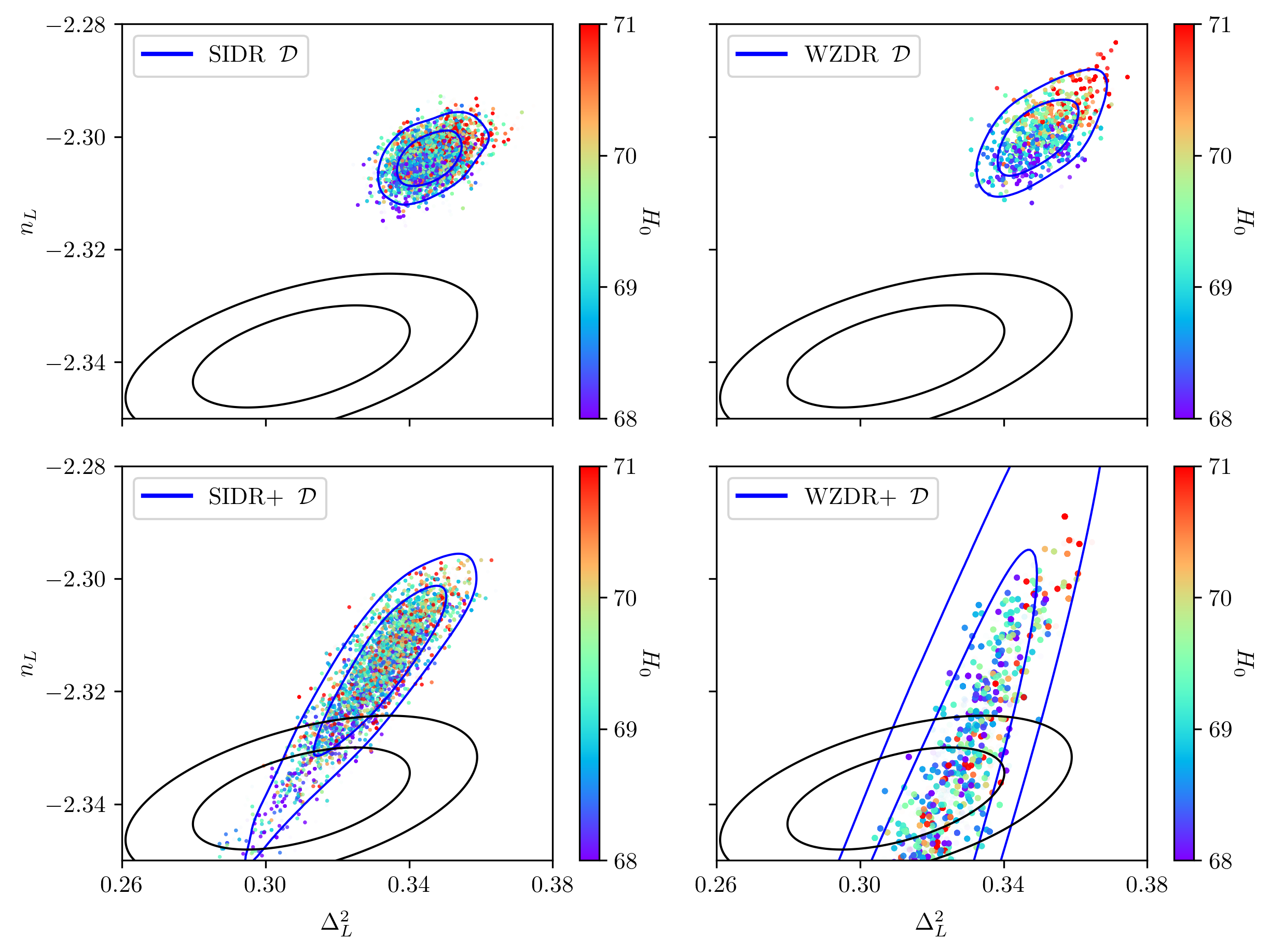}
    \caption{\textbf{(Top Row)} These panels show parameters similar to Fig.~\ref{fig:1}, adjusted for SIDR and WZDR models, where the tension between eBOSS \(\text{Ly}\alpha\) and \Ho remains. \textbf{(Bottom Row)} Similar plots for SIDR+ and WZDR+ models, which fit within the eBOSS parameter space and are thus compatible with \(\text{\dataDL}\) and \(\text{\dataDHL}\) datasets. For these models, there is no clear correlation between \Ho measurements and $\Delta^2_L$ or $n_L$.}
    \label{fig:3}
\end{figure*}

In this section, we  present our full analysis of the extra fluid models SIDR and WZDR, especially focusing on the DM-DR interaction variants, SIDR+ and WZDR+. 

The immediate question that arises is whether the inclusion of eBOSS \lya data is compatible with any of these four models. To address this, we fit our models to the \dataD dataset and plot the posteriors for the \lya observables $\Delta^2_L$ and $n_L$ in Fig.~\ref{fig:3}. We find that, similar to the situation with \lcdm and \Neff models in Fig.~\ref{fig:1}, the fit of SIDR and WZDR models to the \dataD dataset is in approximately 5$\sigma$ tension with eBOSS \lya data. Both SIDR and WZDR struggle to accommodate \lya data; in fact, as they attempt to match steeper slopes for the MPS while still fitting the Planck CMB data, their prediction of \Ho decreases, exacerbating the tension with SH0ES. Fig.~\ref{fig:3} further shows that the correlation between \Ho and \lya is stronger in the WZDR model than in SIDR. This suggests more tension in WZDR if a combined fit to \dataDHL is considered.

A naive interpretation might suggest that the \lya measurement challenges the SH0ES \Ho values. However, the significant variance between the two different \lya likelihoods (eBOSS and XQ-100) possibly points to an unaccounted systematic error in one of them, making reliable interpretation challenging without identifying the error source. To mitigate the impact of systematics, one approach, as Hill et al.~\cite{Goldstein:2023gnw} pursued, is to expand the experimental error margins. A second approach is to take all data at face value and assume that the discrepancies between measurements are due to statistical fluctuations. Given the $\sim 5\sigma$ level of discrepancy, this probability is extremely low. Consequently, we do not prioritize combining fits from all measurements, including XQ-100 of \lya, to achieve the highest possible accuracy in our statistical analysis. However, for the interested reader, we have included triangle plots that combine SH0ES and eBOSS \lya data for the SIDR and WZDR models in Appendix~\ref{app:triangle_plts}. The final possibility, and the primary focus of this paper, is that the \lya measurements, as well as the \dataD and \Ho data, are correct, and that the discrepancy indicates the presence of new physics not explained by SIDR or WZDR models.

As previously motivated in Section~\ref{subsec:models_summary}, this new physics may be a coupling between the DM and DR. Fig.~\ref{fig:3} (bottom row) illustrates how incorporating the parameter $R_{\Gamma}$ in these models expands the parameter space to accommodate eBOSS Ly$\alpha$ measurements. Moreover, the color scatter plot clearly shows the disappearance of the correlation with \Ho predictions, suggesting that the WZDR+ and SIDR+ models can simultaneously provide satisfactory fits to both \Ho and eBOSS \lya.

\begin{figure}[!ht]
    \includegraphics[width=0.49\textwidth]{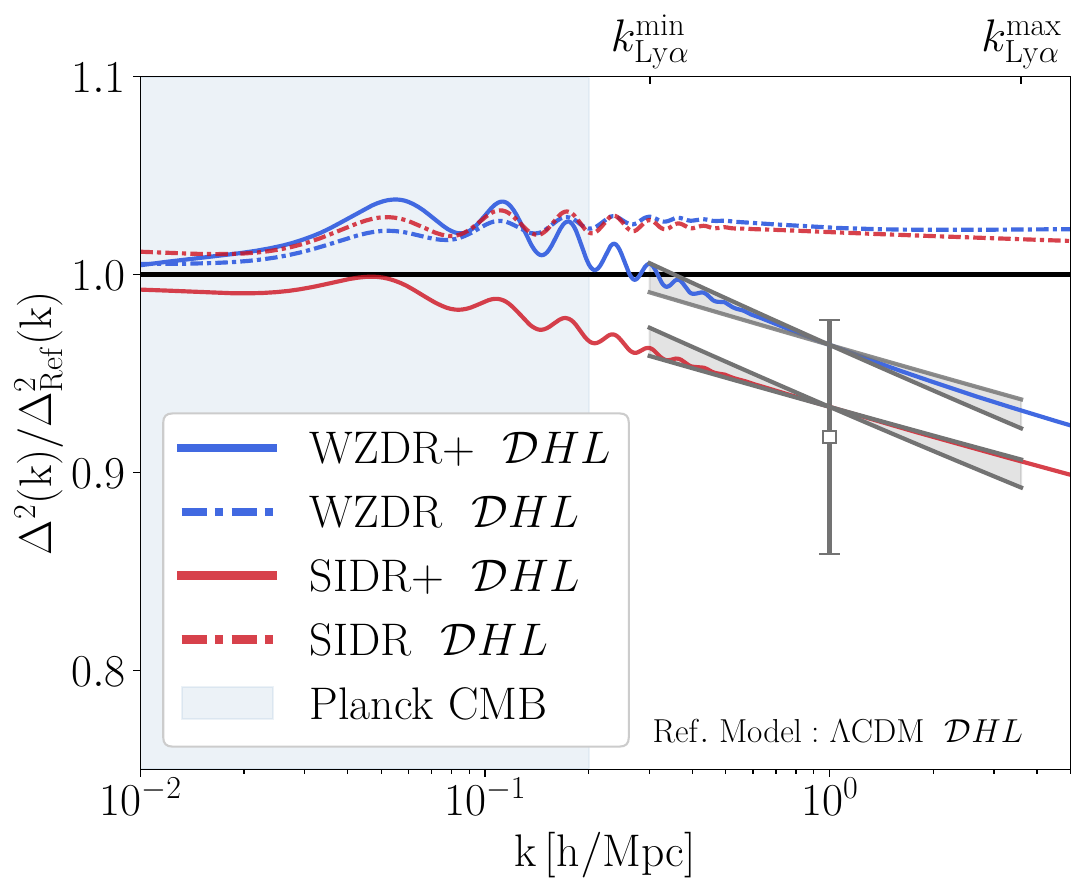}
    \caption{The plot shows the best-fit matter power spectra ratios of selected models to \lcdm for the \dataDHL dataset at $z=0$. The blue shaded area indicates the wavenumbers to which Planck CMB data is sensitive. A white square marks the linear eBOSS \lya data at $k\approx 1\, h/\Mpc$, with gray error bars for the $1\sigma$ amplitude uncertainty $\Delta^2_L$. Tilted cones show the $1\sigma$ uncertainty in $n_L$ of \lya, within the $k_{\mathrm{Ly}\alpha}^{\rm min}<k<k_{\mathrm{Ly}\alpha}^{\rm max}$ range. Adding an interaction strength parameter significantly improves the fit for SIDR+ and WZDR+ models over SIDR and WZDR, highlighting the tension of models without DM-DR interaction with \lya data.}\label{fig:4}
\end{figure}

Another visual depiction of the \lya tension against \lcdm, SIDR, and WZDR, along with the resolution offered by the WZDR+ and SIDR+ models, is illustrated in Fig.~\ref{fig:4}. With the appropriate interaction strength, these + models smoothly dampen the MPS tail to align with the \lya slope.

\begin{figure}[!ht]
    \centering
    \includegraphics[width=0.49\textwidth]{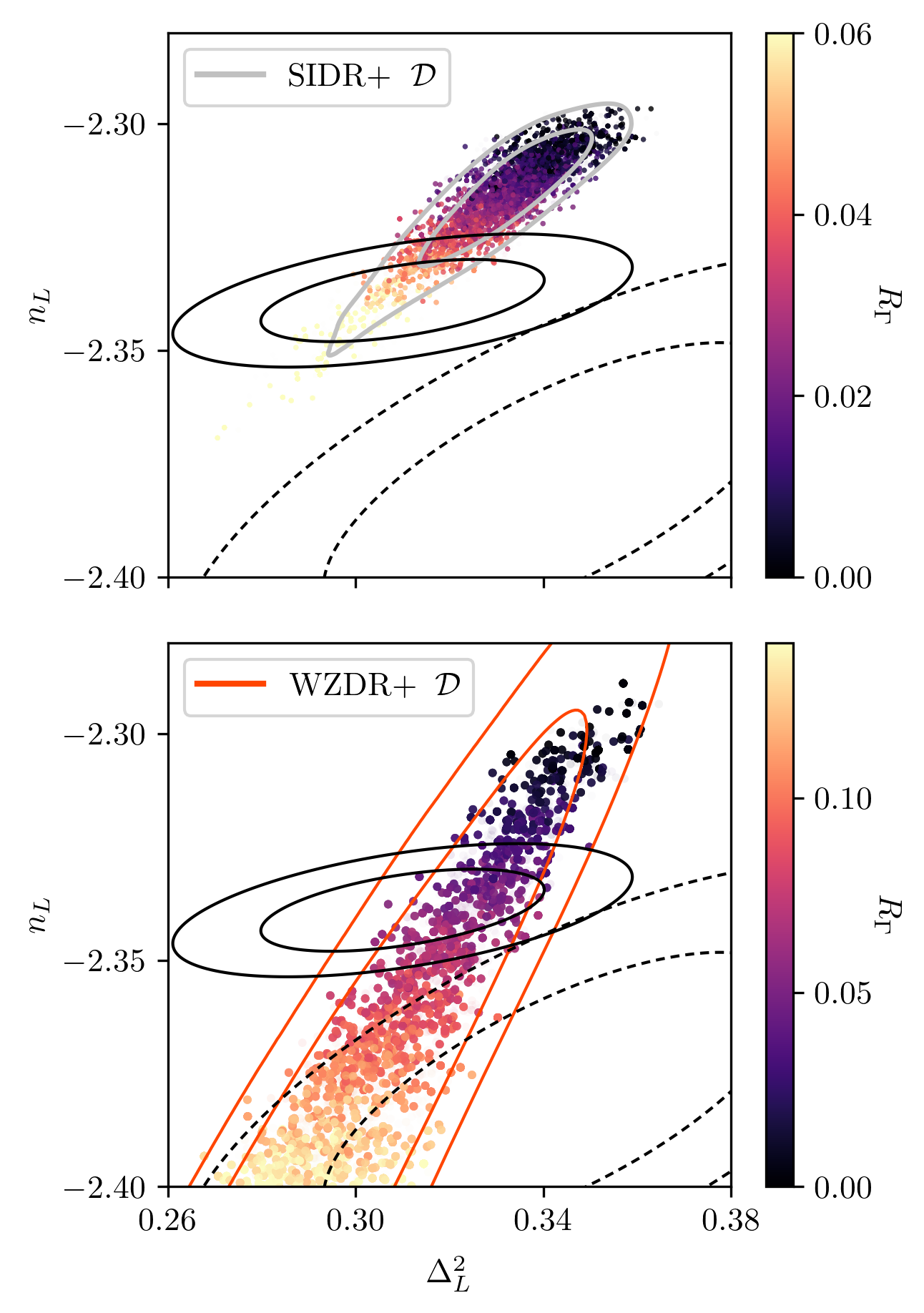}
    \caption{\textbf{(Top)} Marginalized posterior distributions for \lya parameters from SIDR+ fits to \dataD (silver contours), with $R_{\Gamma}$ shown as a color gradient. Solid and dashed black contours indicate $68\%$ and $95\%$ confidence regions from eBOSS and XQ-100 \lya data, respectively (see Appendix~\ref{app:Lya_deets}). The analysis suggests eBOSS data alignment requires non-zero $R_{\Gamma}$, with SIDR+ struggling to fit XQ-100 data. \textbf{(Bottom)} Similarly for WZDR+. Note that the transition redshift $z_t$ opened up a degeneracy direction with increasing $z_t$ and $R_\Gamma$ that allows good matches to either eBOSS or XQ-100 data. This degeneracy is fixed by including eBOSS data into the analysis likelihood (see Fig.~\ref{fig:6}).}\label{fig:5} 
\end{figure}

Fig.~\ref{fig:5} (Top) presents the 2D posteriors of the \lya power spectrum amplitude $\Delta_L^2$ and slope $n_L$ for an SIDR+ model fit to the \dataD dataset, demonstrating how a larger momentum transfer ratio $R_{\Gamma}$ aligns well with the \lya data and remains within the 2D posterior for a fit that includes Planck CMB. As $R_{\Gamma}$ increases, the fit points move towards the solid black Gaussian regions of the \lya contours, reflecting the reduction of power at small scales to match $n_L$.

Fig.~\ref{fig:5} (Bottom) shows that the WZDR+ model can easily fit either of the \lya measurements (eBOSS and XQ-100) along with the Planck CMB data from \dataD. By introducing the $z_t$ parameter, it allows for a break in the MPS at larger $k_t$ creating a flat direction in the parameter space. \lya is the only probe sensitive at small enough scales to resolve this degeneracy, enabling the model to match either measurement by adjusting $z_t$ and $R_{\Gamma}$, as depicted in Fig.~\ref{fig:5} (Bottom).

Meanwhile, within the SIDR+ model, which breaks the MPS around $k_{\rm eq}$, fitting the central values of XQ-100's higher $\Delta^2_L$ and steeper $n_L$ significantly degrades the CMB and full-shape fits. Therefore, SIDR+ can only accommodate the central values of the eBOSS measurement.

To quantify this statement and determine the data's preference for one model over another, we calculate the Gaussian Tensions (GT) in Table~\ref{tb:1}. GT measures the difference in the mean values of the posteriors in units of total $\sigma$, which is the quadrature sum of each standard deviation. The ability of WZDR+ to fit either of the \lya datasets is now evident in Table~\ref{tb:1}, where WZDR+ shows no tension with \lya data in the case of both eBOSS and XQ-100 likelihoods, while a $\sim 4 \sigma$ tension persists for SIDR+ with the XQ-100 measurement. 

\newcolumntype{L}{>{\centering\arraybackslash}X}
\begin{table}[t!]
    \setlength\extrarowheight{3pt}
    \centering
    \begin{tabularx}{0.49\textwidth}{|c | c | c | c | c | L| } 
        \hline
        & \lcdm & WZDR+ & SIDR+ &WZDR & SIDR\\
        \hline
        $GT$ - eBOSS & $4.9 \sigma$  & $0.58\sigma$  & $1.3\sigma$ & $5.0 \sigma$ & $4.9 \sigma$ \\  \hline
        
        $GT$ - XQ-100 & $4.5 \sigma$  & $1.0\sigma$  & $3.9\sigma$ & $4.6 \sigma$ & $4.5 \sigma$ \\  \hline
    \end{tabularx}
    \caption{Gaussian Tension between direct measurements of \lya and from \lcdm, WZDR+, SIDR+, WZDR and SIDR fit to \dataD.}
    \label{tb:1}
\end{table}

Models WZDR and SIDR, without the DM-DR interactions, fail to suppress the MPS at \lya scales while fitting the CMB. Fitting these to \dataDL and comparing to \dataD, we observe a $\Delta \chi^2_{\rm CMB}\sim 7$, while still being in strong disagreement with eBOSS \lya. The full set of $\chi^2$ values are provided for the interested reader in Table~\ref{tb:8} in Appendix~\ref{app:triangle_plts}. For SIDR+ and WZDR+, $\Delta \chi^2_{\rm CMB}$ improves to $\sim5.6$ and $\sim0.4$, respectively. SIDR+ worsens the CMB $\chi^2$ in a combined fit that includes \lya, by suppressing the MPS near $k\sim k_{\rm eq}$, to which the CMB data are still sensitive. The WZDR+ model avoids this problem by shifting the break in the MPS to smaller scales. Both + models significantly improve the \lya fit over their non-interacting counterparts.

\begin{figure*}[t]
    \centering
    \begin{minipage}{0.49\textwidth}
        \centering
        \includegraphics[width=\textwidth]{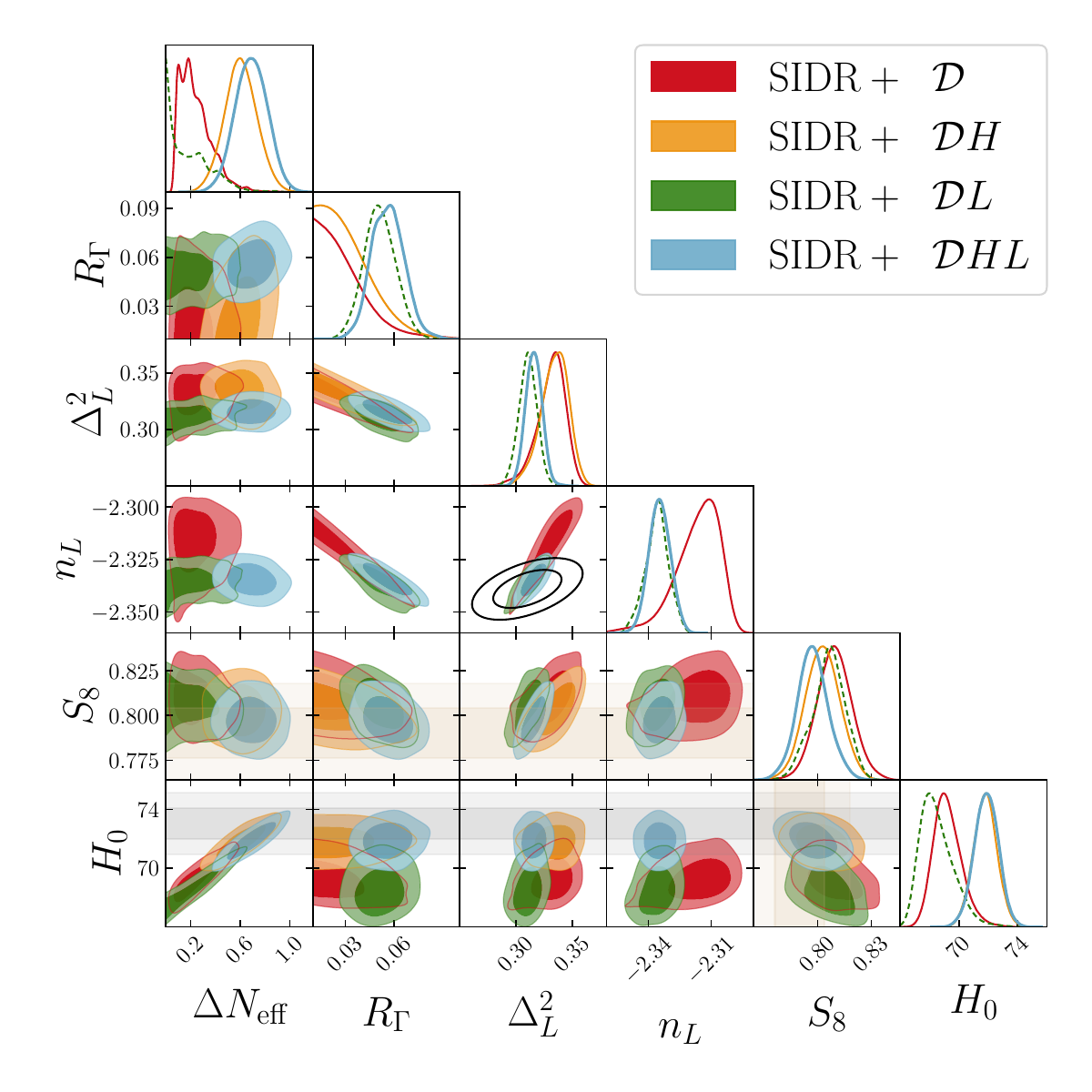}
    \end{minipage}\hfill
    \begin{minipage}{0.49\textwidth}
    \centering
        \includegraphics[width=\textwidth]{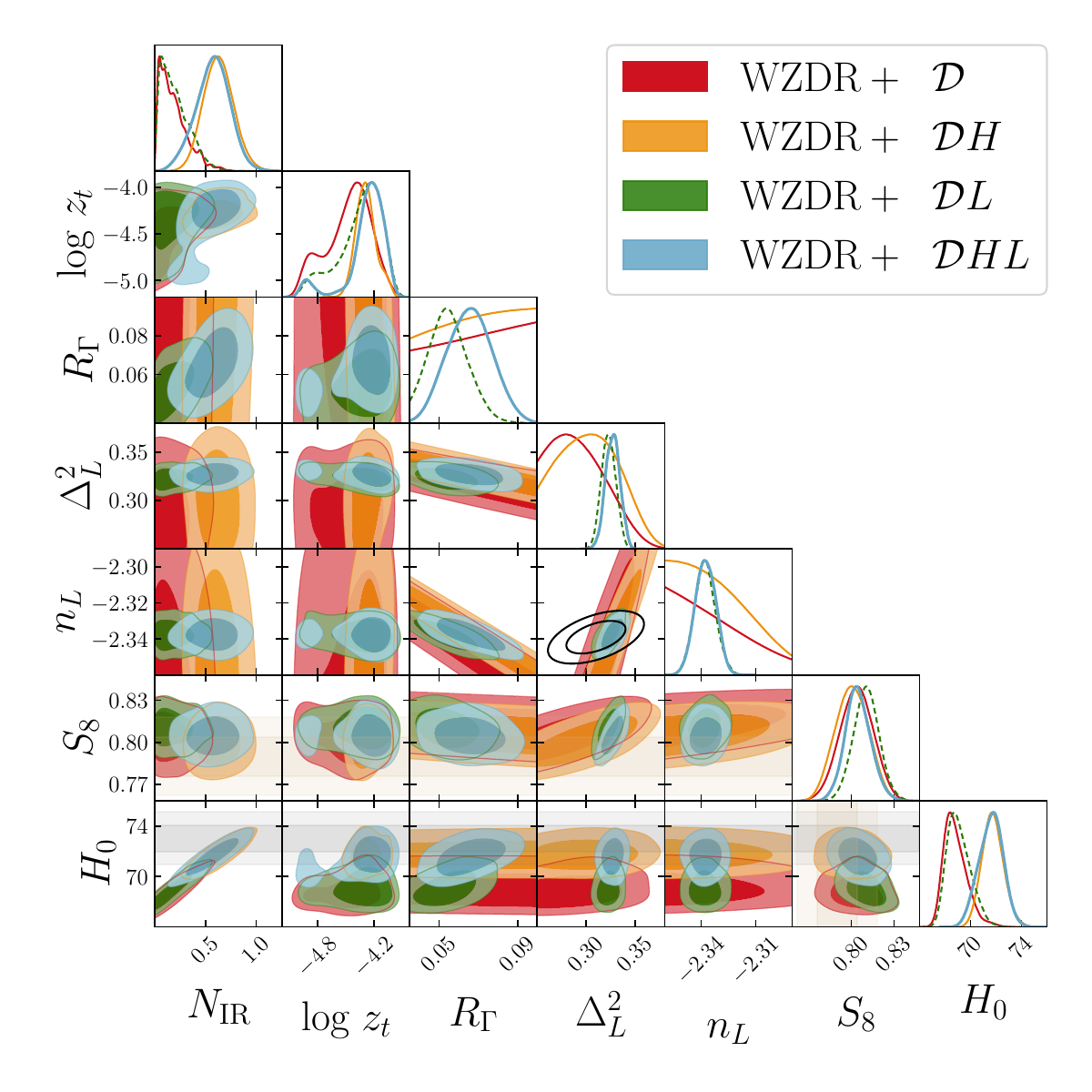}
    \end{minipage}
    \caption{\textbf{(Left)} Triangle plot of additional cosmological parameters for the SIDR+ model, along with SIDR+'s predictions for the \lya slope and amplitude, and \Ho. The grey bands display the latest SH0ES measurement, and the solid black ellipses correspond to the $68\%$ and $95\%$ confidence levels from the eBOSS \lya measurement. \textbf{(Right)} The same plot for the WZDR+ model.
}\label{fig:6}
\end{figure*}

We also compare SIDR+ and WZDR+ model parameter posteriors across four datasets in Fig.\ref{fig:6}. The \Ho - \nir panels show a clear preference for non-zero \nir in \dataDH and \dataDHL to match SH0ES measurements. Since \nir and $R_{\Gamma}$ are uncorrelated, adjusting $R_{\Gamma}$ to fit \dataDL and \dataDHL is straightforward. The inclusion of $S_8$ and MPS data in \dataD indicates a preference for non-zero $R_{\Gamma}$ (see~\Cref{tb:2,tb:3}). Furthermore, a preference for $\mathrm{log}\,z_t\sim -4.2$ in WZDR+ across all datasets, as shown in Table~\ref{tb:3} and discussed in previous works~\cite{Aloni:2021eaq,Joseph:2022jsf}, may suggest that new physics is active at this scale. This preference, potentially due to adjustments in CMB phase shifts and the break in the MPS, is remarkably consistent across datasets, with \dataD values staying within $1\sigma$ of those from \dataDL or \dataDHL.

\newcolumntype{L}{>{\centering\arraybackslash}X}

\begin{table}[!ht]
\centering
\begin{minipage}{0.49\textwidth}
\centering
    \begin{tabularx}{\textwidth}{|L|c|L|L|c|}
        \hline
        $R_{\Gamma}$ & \dataD                         & \dataDH                  & \dataDL                  & \dataDHL                    \\ \hline
        Bestfit      & 0.0032                         & 0.0082                   & 0.052                    & 0.055                       \\ \hline
        Mean         & $\,0.0226_{-0.0054~~}^{+0.0226~~}$ & $\,0.03_{-0.01}^{+0.03}$ & $\,0.05_{-0.01}^{+0.01}$ & $\,0.056_{-0.009}^{+0.011}$ \\ \hline
    \end{tabularx}
    \caption{Interaction strength parameter for SIDR+ in different datasets.}
    \label{tb:2}
\end{minipage}\vskip0.25cm\hfill
\begin{minipage}{0.49\textwidth}
\centering
    \begin{tabularx}{\textwidth}{LLLLL}
        \hline
        \multicolumn{1}{|c|}{$\mathrm{log}~z_t$} & \multicolumn{1}{c|}{\dataD}                  & \multicolumn{1}{c|}{\dataDH}                 & \multicolumn{1}{c|}{\dataDL}                   & \multicolumn{1}{c|}{\dataDHL}                \\ \hline
        \multicolumn{1}{|c|}{Bestfit}            & \multicolumn{1}{c|}{-4.20}                   & \multicolumn{1}{c|}{-4.29}                   & \multicolumn{1}{c|}{-4.20}                     & \multicolumn{1}{c|}{-4.26}                   \\ \hline
        \multicolumn{1}{|c|}{Mean}               & \multicolumn{1}{c|}{-$4.46_{-0.19}^{+0.29}$} & \multicolumn{1}{c|}{-$4.29_{-0.12}^{+0.09}$} & \multicolumn{1}{c|}{-$4.33_{-0.07}^{+0.33}$}   & \multicolumn{1}{c|}{-$4.25_{-0.12}^{+0.15}$} \\ \hline
                                                 &                                              &                                              &                                                &                                              \\ \hline
        \multicolumn{1}{|c|}{$R_{\Gamma}$}       & \multicolumn{1}{c|}{\dataD}                  & \multicolumn{1}{c|}{\dataDH}                 & \multicolumn{1}{c|}{\dataDL}                   & \multicolumn{1}{c|}{\dataDHL}                \\ \hline
        \multicolumn{1}{|c|}{Bestfit}            & \multicolumn{1}{c|}{0.067}                   & \multicolumn{1}{c|}{0.11}                    & \multicolumn{1}{c|}{0.052}                     & \multicolumn{1}{c|}{0.070}                   \\ \hline
        \multicolumn{1}{|c|}{Mean}               & \multicolumn{1}{c|}{$0.22_{-0.22}^{+0.04}$}  & \multicolumn{1}{c|}{$0.15_{-0.11}^{+0.07}$}  & \multicolumn{1}{c|}{$0.054_{-0.011}^{+0.010}$} & \multicolumn{1}{c|}{$0.07_{-0.01}^{+0.01}$}  \\ \hline
    \end{tabularx}%
    \caption{Step and interaction strength parameters for WZDR+ in different datasets.}
    \label{tb:3}
\end{minipage}
\end{table}

\subsection{$S_8$ and full-shape analysis}

An analysis of the full-shape data and its constraining power on the WZDR+ model has been conducted in~\cite{Schoneberg:2023rnx,Allali:2023zbi}. The main finding by Schöneberg et al. was that the full-shape data imposes stronger constraints on the allowed interaction strength $R_{\Gamma}$, when combined with a measurement of $S_8 = 0.769^{+0.016}_{-0.012}$~\cite{Heymans:2020gsg,DES:2021wwk} that is in strong tension with Planck. This result suggests that the full-shape data disfavors stronger interaction strengths or, correspondingly, steeper slopes that arise from their chosen $S_8$ likelihood, resulting in an upward shift of $+0.7\sigma$ on the $S_8$ tension. Our analysis is consistent with their findings; however, with the updated cosmic shear analysis of $S_8 = 0.790^{+0.018}_{-0.014}$~\cite{Kilo-DegreeSurvey:2023gfr}, there is no tension with the full-shape data in the DM-DR interacting models, as indicated by the $\chi^2$ values in Table~\ref{tb:8}.

 \section{Discussion}\label{sec:conclusions}

Models which contain dark radiation beyond the Standard Model neutrinos predict a shorter ``standard ruler", i.e. sound horizon, as a solution to the Hubble tension. We compare such models to LSS data, and find that all the models considered, \lcdm, SIDR, WZDR, SIDR+, and WZDR+ are in good agreement with CMB lensing, weak lensing, and the full-shape galaxy power spectrum. However, the situation with \lya data is more complex.

The \lya forest is unique in that it probes shorter co-moving length scales than the CMB and other LSS data. It is sensitive to physical processes that acted at an earlier time in the history of the universe, when the wave modes measured in the CMB were still "frozen" outside the horizon. Therefore, \lya data are particularly powerful in constraining models that impact the growth of structure at early times during radiation domination. This is the period where so-called "early universe" solutions to the Hubble tension introduce new physics. Interestingly, and perhaps relatedly, the eBOSS 1D \lya flux power spectrum measurements are also in $\sim 5\sigma$ tension with \lcdm predictions when fit to the CMB~\cite{Rogers:2023upm}.~\cite{Goldstein:2023gnw} found that this tension becomes worse in models with EDE designed to solve the Hubble tension. 

We find a similar situation for the SIDR and WZDR dark radiation models. Their predictions for the MPS at $\sim h \mathrm{Mpc}^{-1}$ scales are very similar to those of \lcdm, but we also find a positive correlation between increasing values of \Ho, aimed at resolving the Hubble tension, and an increase of the amplitude and slope of the MPS. Thus in the region of parameter space where the SIDR and WZDR models significantly improve the Hubble tension they worsen the fit to eBOSS \lya data. 

The tension between eBOSS \lya data and the predictions from \lcdm, SIDR, and WZDR fit to CMB and BAO data allows three possible interpretations: {\it i.} The \lya data or its interpretation as a measurement of the linear MPS may have a significant, hitherto unidentified systematic error, which is at the root of the tension. If this is the case, we do not have much to say, since we do not know either the magnitude or direction of this ``unknown unknown." {\it ii.} It is possible, although highly unlikely, that the $\sim 5\sigma$ disagreement is the result of a statistical fluctuation. In that case, the correlation of higher values of \Ho with an increased MPS worsens the Hubble tension within \lcdm and also strongly disfavors SIDR and WZDR as solutions to the tension. {\it iii.} If the tension is neither due to a statistical fluctuation nor a systematic error, it calls for new physics that suppresses the MPS at small scales. SIDR+ and WZDR+ are two possible models in which such a suppression naturally occurs due to the scattering of dark radiation off dark matter. In particular, as was first shown in~\cite{Buen-Abad:2015ova}, if the momentum transfer rate between dark matter and dark fluid scales proportional to $T^2$, like the Hubble expansion rate, then a smooth suppression of the MPS at small scales results. This suppression takes the form of a break in the power of the k-dependence: 
\begin{equation}
\label{eq:slopediscussion}
    \Delta^2_{R_{\Gamma}\neq 0}\simeq \begin{cases}
        \Delta^2_{R_{\Gamma}= 0} \quad &k \ll k_{\rm break}\\
        \Delta^2_{R_{\Gamma}= 0}\left(\frac{k}{k_{\rm break}}\right)^{-R_{\Gamma}} \quad &k \gg k_{\rm break}
    \end{cases}\,,
\end{equation}
where the exponent $R_{\Gamma}$ is related to the strength of DM-DR scattering, and $k_{\rm break}$ is either $k_{\rm eq}$ in the case of SIDR+ or the larger transition wave number $k_t$ in WZDR+. A good fit to all data, including the \lya forest, requires $R_{\Gamma} \sim 0.05$.

Given the discrepancy between the measurements from eBOSS and XQ-100, it may be premature to claim evidence for new physics. See also~\cite{Fernandez:2023grg} for a recent reanalysis of eBOSS data based on new simulations, which points to a tension between different redshift bins of the eBOSS data.
We therefore look forward to the release of results from the \lya 1D flux power spectrum measured from the DESI year 1 dataset, which is currently being analyzed. The increased resolution of the DESI spectrograph, along with new synthetic data from improved simulations and careful analysis as demonstrated with the DESI Early Data Release~\cite{DESI:2023xwh,Karacayli:2023afs}, promises to make this a particularly exciting result. It will not only directly improve our knowledge of the MPS at $h\mathrm{Mpc}^{-1}$ scales but also shed indirect light on the Hubble tension by distinguishing between ``early universe models."
As we showed in this paper, models with a DM-DR interaction such as SIDR+ and WZDR+ provide an excellent fit to CMB, BAO, uncalibrated supernova, and large-scale structure data including \lya, while slicing the Hubble tension in half.

\section{Notes added}
As this work was being prepared the DESI BAO measurement~\cite{DESI:2024mwx} was announced, finding $\Omega_M=0.295\pm0.015$ in \lcdm. As this is in excellent agreement with our fits (See Tables~\ref{tb:5},~\ref{tb:6},and ~\ref{tb:7} in Appendix~\ref{app:triangle_plts}) and the models we studied do not modify the late time expansion history we do not expect this measurement to significantly impact our results. In addition, an EFTofLSS analysis of the BOSS and eBOSS \lya data~\cite{Ivanov:2024jtl} appeared. This work conservatively focuses on a subset of the data and includes a number of bias parameters to reduce astrophysical model dependence.~\cite{Ivanov:2024jtl} perform a fit to \lcdm with all parameters except $\sigma_8$ fixed from a Planck2018 and determines $\sigma_8=0.841\pm0.017$, consistent with Planck and higher than other direct LSS probes. We also find agreement with \lcdm when fitting only the overall amplitude (and not the slope) of the MPS with the $\Delta_L^2+n_L$ likelihood. It would be interesting to see if the EFT of \lya can be used to determine the more model-independent slope and amplitude of the linear MPS at $h\mathrm{Mpc}^{-1}$.

\vskip0.5cm
{\bf Acknowledgments}\quad
We thank Andreu Font Ribera and Satya Gontcho A Gontcho for useful discussions on the validity of approaches to fit the eBOSS \lya data. H.B. is supported by the DOE Grant DE-SC0013607 and NASA Grant 80NSSC20K0506. The work of M.S. and E.N.S. is supported by DOE under Award DE-SC0015845. M.S. thanks the CCPP at NYU and the Physics Department at Göttingen University for their hospitality and support. Our MCMC runs were performed on the Shared Computing Cluster, which is administered by Boston University’s Research Computing Services.

\bibliography{lya_NODUP2.bib}
\appendix


\section{Details of \lya likelihood}\label{app:Lya_deets}

\begin{figure}[h!]
    \centering
    \begin{minipage}{0.49\textwidth}
        \centering
        \includegraphics[width=\textwidth]{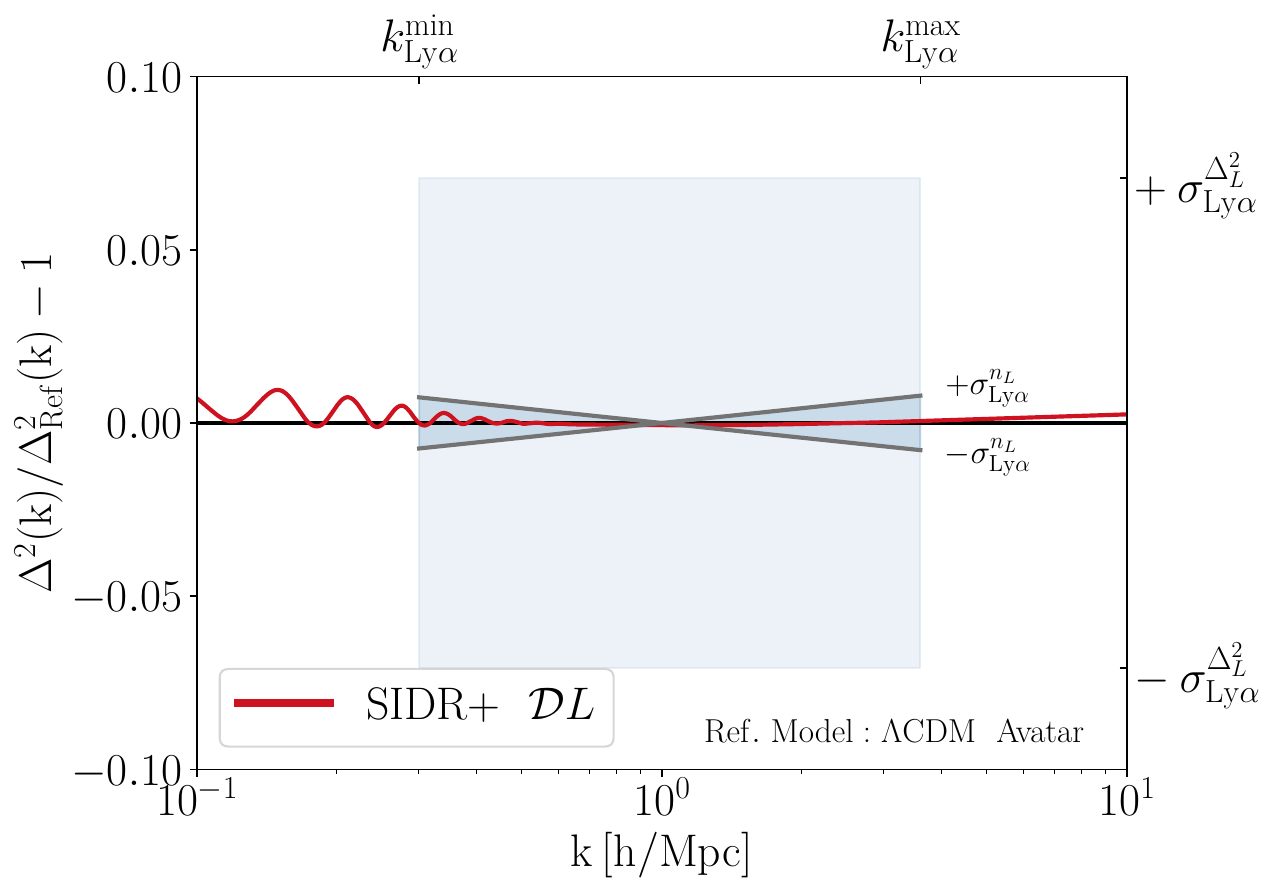}
    \end{minipage}\hfill
    \begin{minipage}{0.49\textwidth}
    \centering
        \includegraphics[width=\textwidth]{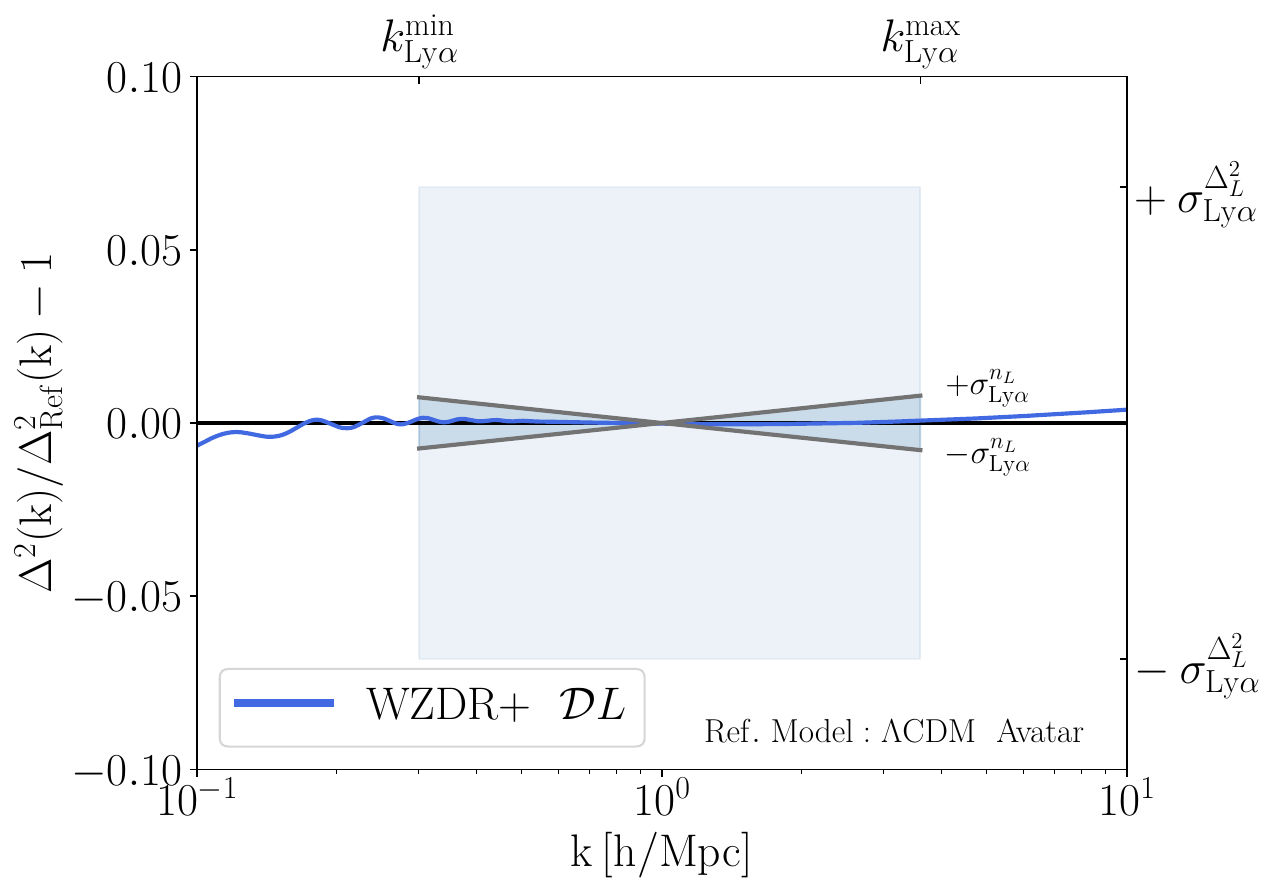}
    \end{minipage}
    \caption{\textbf{(Top)}  Ratio of the MPS for the SIDR+ best fit to the \dataDL dataset compared to the corresponding \lcdm avatar, with parameters detailed in Table~\ref{tb:4}. The shaded cones and the height of the shaded rectangle represent the errors in the compressed eBOSS \lya measurement, normalized to the underlying \lcdm avatar. The width of the shaded rectangle indicates the scales included in the measurement. \textbf{(Bottom)} A similar plot for the WZDR+ model.}\label{fig:7} 
\end{figure}

In this Appendix, we detail the derivation of the compressed eBOSS \lya likelihood and address concerns regarding its applicability to \lcdm extensions, specifically SIDR+ and WZDR+. Note that the shape of our MPS suppression does not fit the parameterizations considered in~\cite{Hooper:2022byl}.

To extract information from data influenced by non-linear physics, the eBOSS collaboration~\cite{eBOSS:2018qyj} matched the observed 1D \lya flux power spectrum along the line of sight up to second order by conducting hydrodynamical simulations within the \lcdm cosmology. This involved varying \Ho, $\Omega_M$, $n_s$, and $\sigma_8$~\cite{Borde:2014xsa}. Additionally, the effects of intergalactic medium (IGM) parameters, such as the temperature at mean density ($T_0$) and the spectral index ($\gamma$) at $z = 3$, which defines the temperature-density relation $T = T_0\delta^{\gamma-1}$, were examined.

Readers may be concerned that extracting linear parameters from experimental data, which is sensitive to full non-linear effects, has been conducted within a \lcdm cosmology framework. However, Pedersen et al.~\cite{Pedersen:2022anu} demonstrate that $\Delta^2_L$ and $n_L$ posteriors are model-independent across a range of models that smoothly modify the MPS at early times without altering the late-time growth of structure from \lcdm.

In Fig.~\ref{fig:7}, we show that, within the measurement's error bounds, the slope and amplitude of the SIDR+ and WZDR+ models can be approximated by a \lcdm model (referred to as \textit{avatars}). Error bounds are depicted as shaded cones for the slope and shaded rectangles for the amplitude, with the widths of the rectangles indicating the $k$ scales measured by eBOSS Ly$\alpha$. The red and blue lines represent the best fits for SIDR+ and WZDR+ to \dataDL, normalized to \lcdm avatars. These avatars only differ from a \dataD best fit in the $A_s$ and $n_s$ values, which are adjusted to match the MPS of SIDR+ and WZDR+ best fits to \dataDL, respectively.

The selected $A_s$ and $n_s$ values for these avatars are detailed in Table~\ref{tb:4}. The agreement of the \lcdm power spectrum with both SIDR+ and WZDR+ models is within $0.1\%$ for the eBOSS Ly$\alpha$ sensitive regions, indicating that WZDR+ and SIDR+ share the same linear MPS as \lcdm at these scales. Additionally, they exhibit identical late universe parameters, such as \Ho and $\omega_m$, to those of the \lcdm avatars.

\begin{table}[h!]
\centering
\begin{tabular}{|c|c|c|}
\hline
                             & $ln10^{10}A_s$ & $n_s$ \\ \hline
\lcdm:~$\mathcal{D}$         & 3.046          & 0.968 \\ \hline
\lcdm:~SIDR+ Avatar & 3.016          & 0.935 \\ \hline
\lcdm:~WZDR+ Avatar & 3.059          & 0.933 \\ \hline
\end{tabular}
\caption{Cosmological parameter values for the \lcdm avatars used in Fig.~\ref{fig:7} that differ from those of a best-fit \lcdm to \dataD.}
\label{tb:4}
\end{table}

Given their shared linear MPS and late universe physics, including $\omega_{\rm cdm}$, $\omega_b$, and \Ho, SIDR+ and WZDR+ undergo the same growth. Consequently, the results from N-body simulations are applicable to both SIDR+ and WZDR+ models as well. 

\section{Triangle Plots and Parameter Values}\label{app:triangle_plts}

\begin{figure*}[h!]
    \centering
    \includegraphics[width=0.98\textwidth]{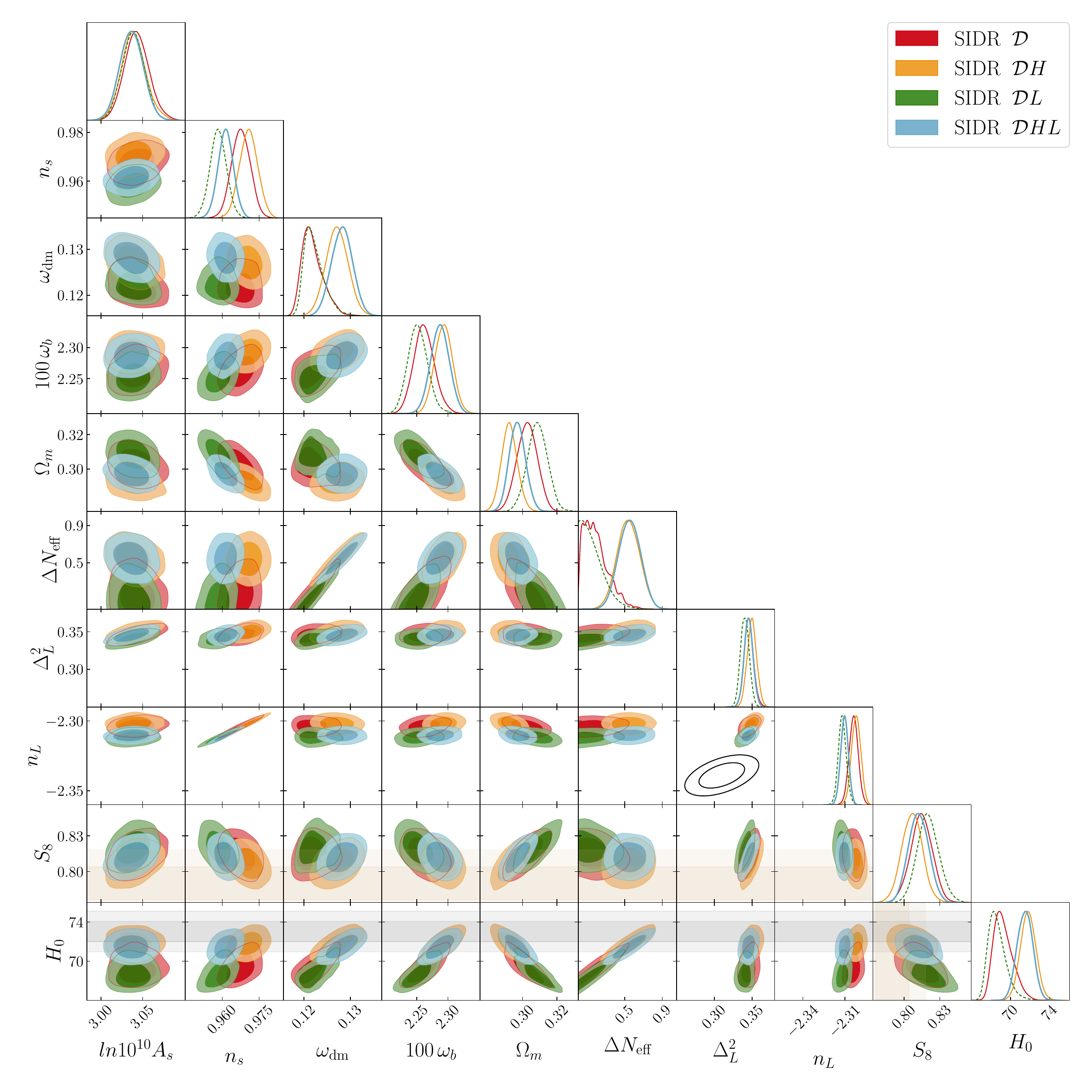}
    \caption{Comparison of SIDR model fits to four datasets, with light and dark shaded regions representing $68\%$ and $95\%$ confidence levels, respectively. The marginalized posteriors for $\Delta_L^2$ and $n_L$ are juxtaposed with the 2D \lya likelihood (black contours).}
    \label{fig:8}
\end{figure*}

\begin{figure*}[h!]
    \centering
    \includegraphics[width=0.98\textwidth]{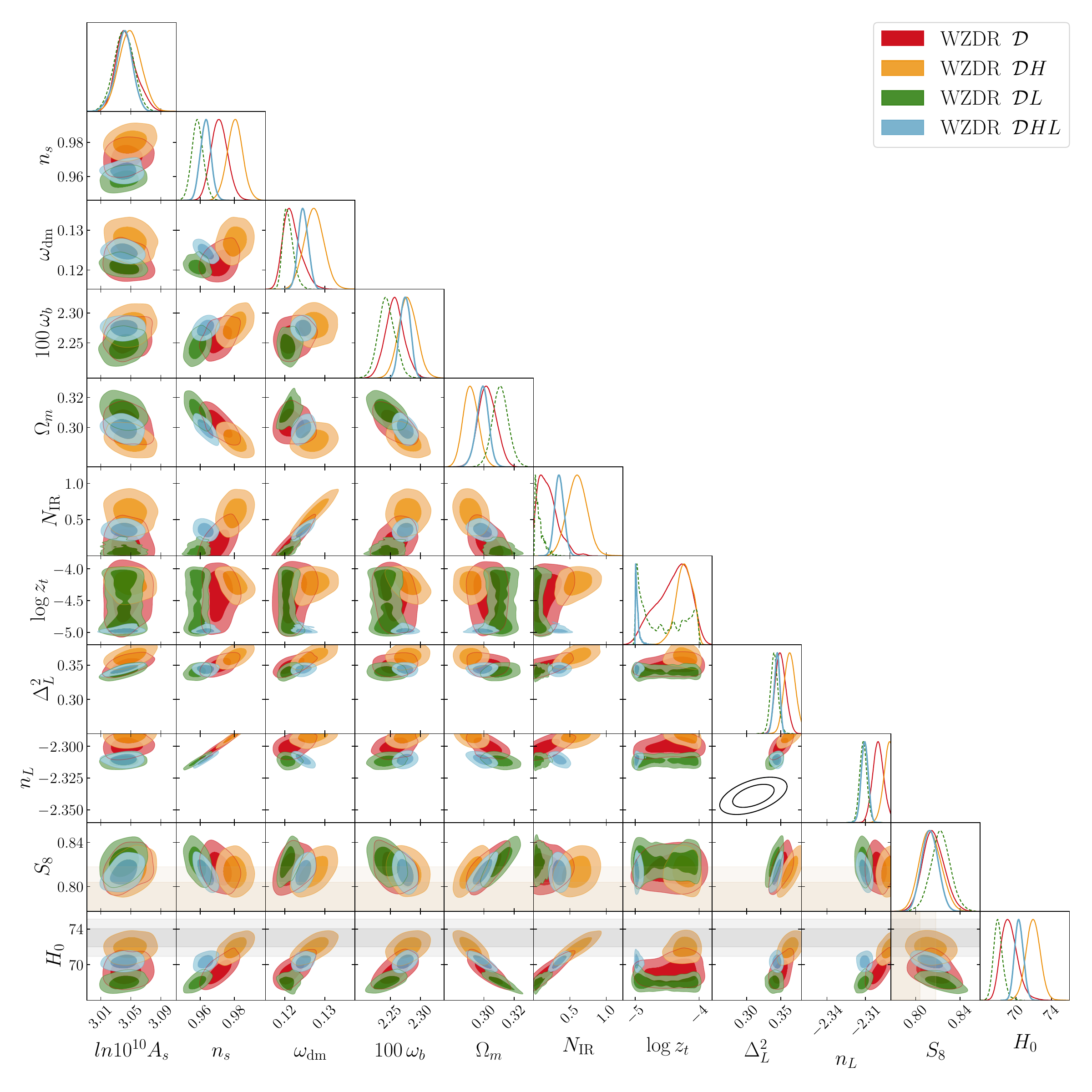}
    \caption{Comparison of WZDR model fits to four datasets, with light and dark shaded regions representing $68\%$ and $95\%$ confidence levels, respectively. The marginalized posteriors for $\Delta_L^2$ and $n_L$ are juxtaposed with the 2D \lya likelihood (black contours).}
    \label{fig:9}
\end{figure*}

\begin{figure*}[h!]
    \centering
    \includegraphics[width=0.98\textwidth]{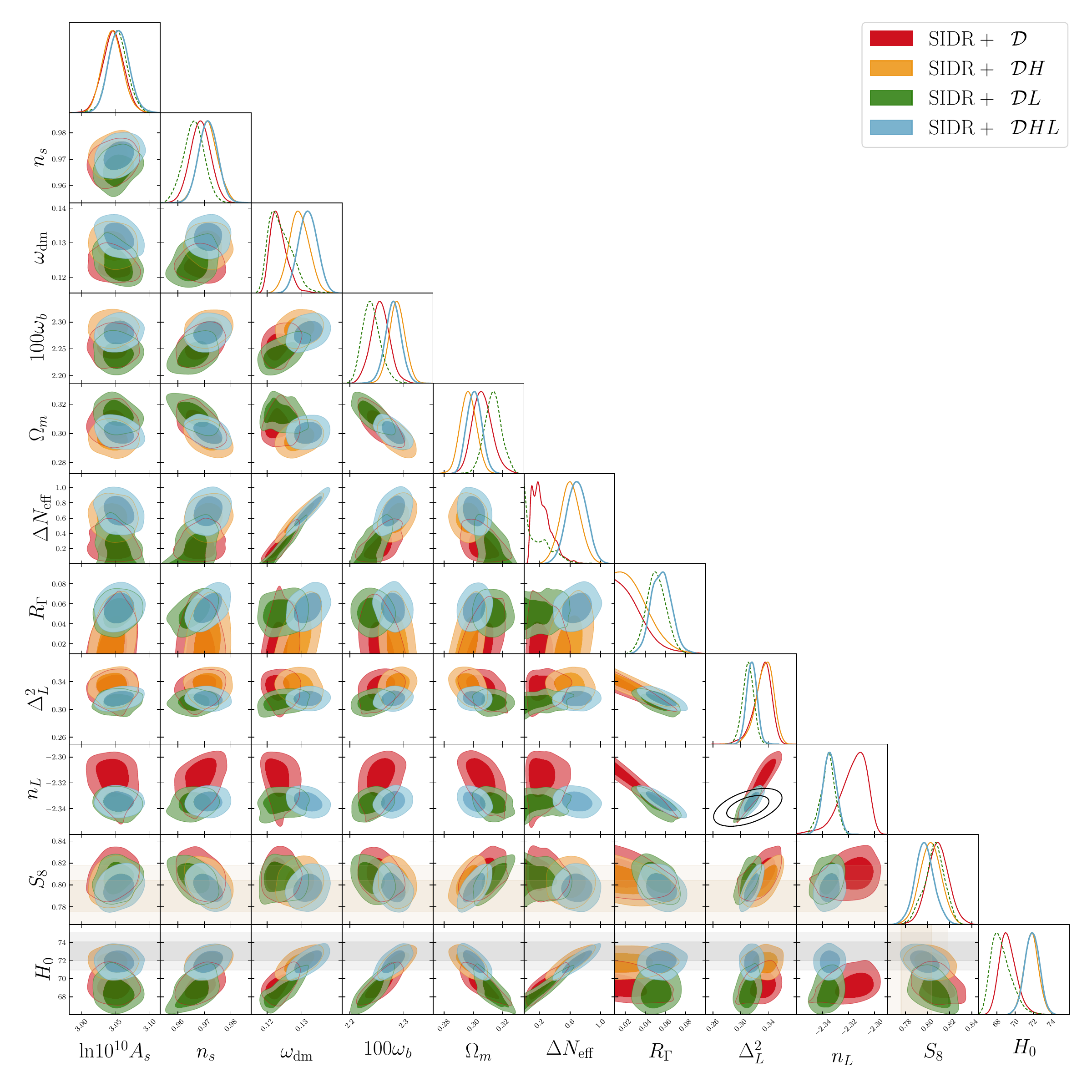}
    \caption{Comparison of SIDR+ model fits to four datasets, with light and dark shaded regions representing $68\%$ and $95\%$ confidence levels, respectively. The marginalized posteriors for $\Delta_L^2$ and $n_L$ are juxtaposed with the 2D \lya likelihood (black contours).}
    \label{fig:10}
\end{figure*}

\begin{figure*}[h!]
	\centering
    \includegraphics[width=0.98\textwidth]{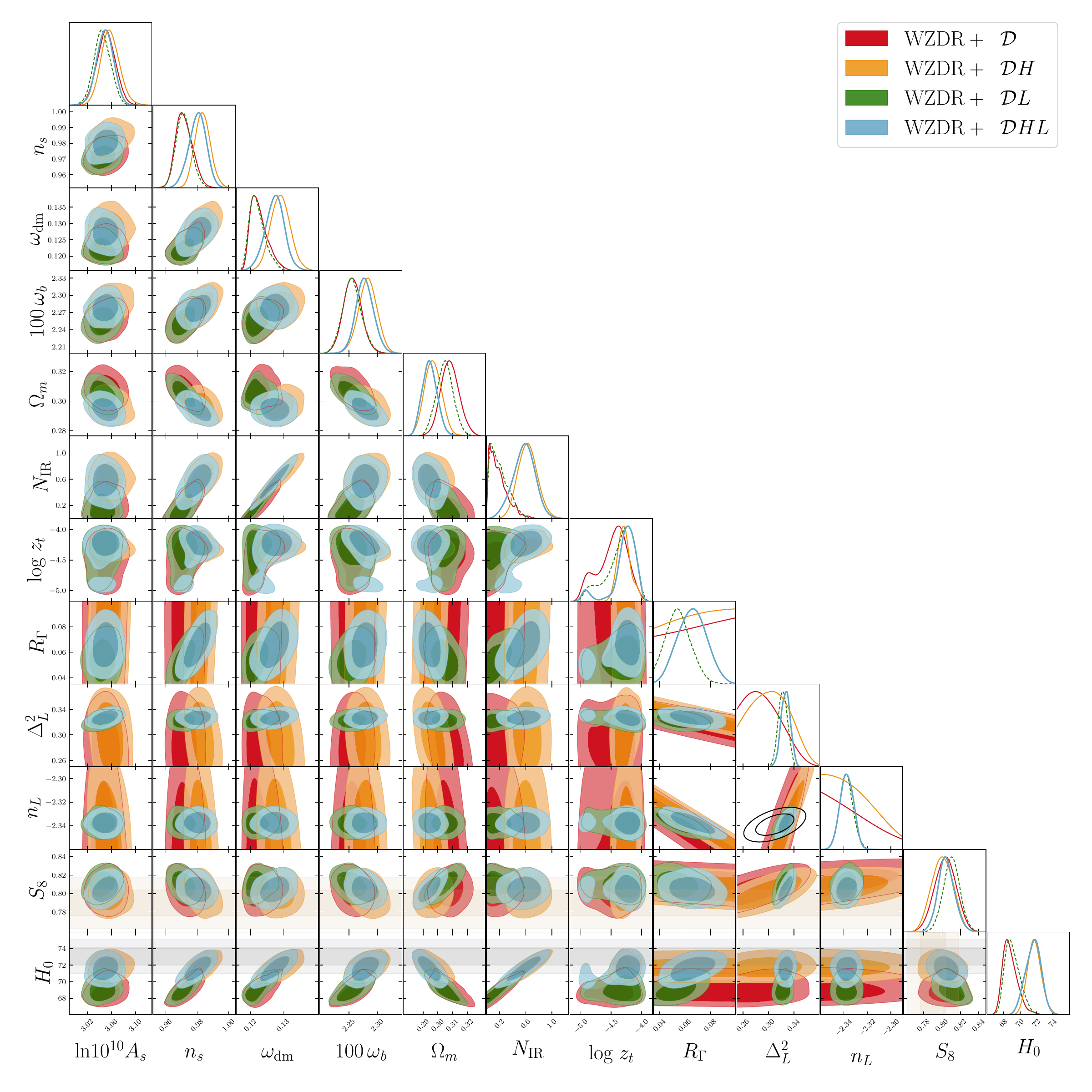}
    \caption{Comparison of WZDR+ model fits to four datasets, with light and dark shaded regions representing $68\%$ and $95\%$ confidence levels, respectively. The marginalized posteriors for $\Delta_L^2$ and $n_L$ are juxtaposed with the 2D \lya likelihood (black contours).}
	\label{fig:11}
\end{figure*}

\begin{figure*}[h!]
	\centering
    \includegraphics[width=0.98\textwidth]{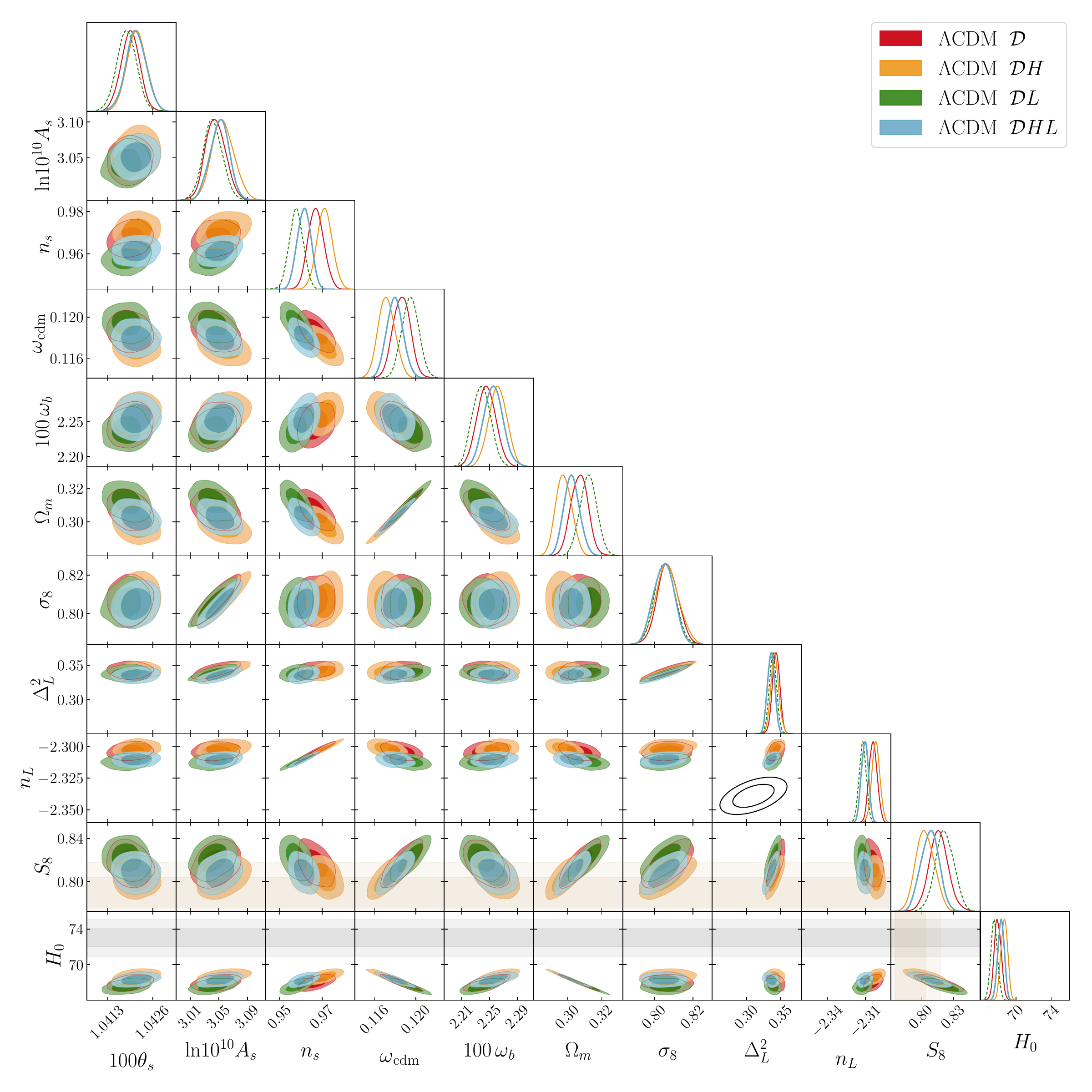}
    \caption{Comparison of \lcdm model fits to four datasets, with light and dark shaded regions representing $68\%$ and $95\%$ confidence levels, respectively. The marginalized posteriors for $\Delta_L^2$ and $n_L$ are juxtaposed with the 2D \lya likelihood (black contours).}
    \label{fig:12}
\end{figure*}

\begin{figure*}[h!]
	\centering
    \includegraphics[width=0.98\textwidth]{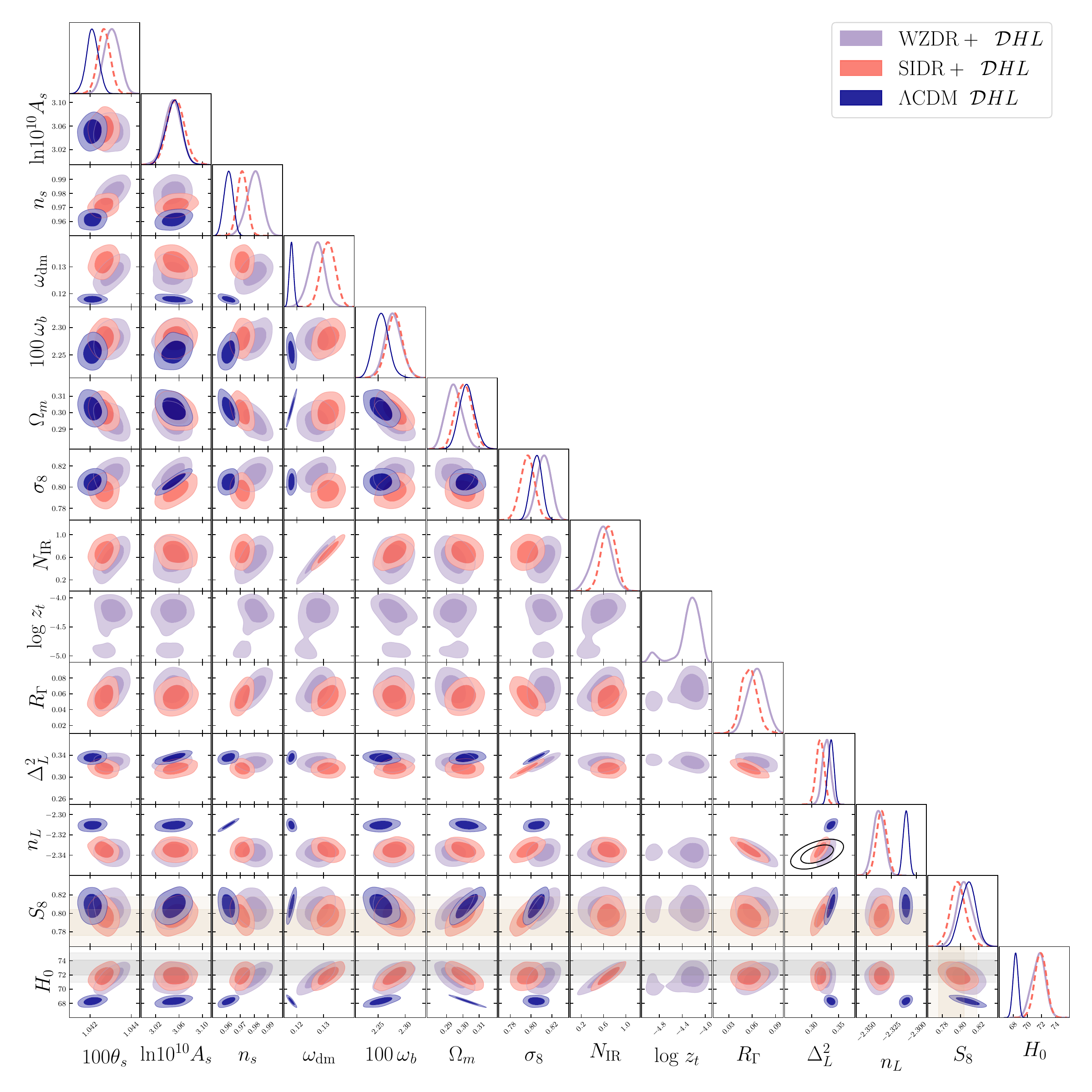}
    \caption{Comparison of \lcdm model parameters from \dataDHL dataset with interacting DM-DR models: WZDR+ and SIDR+, where $N_\mathrm{IR} = \Delta N_\mathrm{eff}$ in SIDR+.}
    \label{fig:13}
\end{figure*}

This Section presents the full set of marginalized posteriors for \lcdm parameters and additional parameters of the models SIDR(+) and WZDR(+). \Cref{fig:8,fig:9,fig:10,fig:11,fig:12} show comparisons of SIDR, WZDR, SIDR+, WZDR+, and \lcdm fits to datasets \dataD, \dataDH, \dataDL, and \dataDHL. Meanwhile,~\ref{fig:13} compares the + model's fits to \lcdm for datasets \dataD, \dataDH, \dataDL, and \dataDHL. 

Tables~\ref{tb:5},~\ref{tb:6}, and~\ref{tb:7} show marginalized 1D posterior means and $68\%$ credible intervals for cosmological parameters across the \lcdm, SIDR+, and WZDR+ models, evaluated for various datasets respectively. Table~\ref{tb:8} incudes the $\chi^2$ for fits of all models to all the datasets.

\begin{table*}[h!]
\centering
\begin{tabular}{c|cc|cc|cc|cc|}
\cline{2-9}
                                            & \multicolumn{2}{c|}{\dataD}                                   & \multicolumn{2}{c|}{\dataDH}                                  & \multicolumn{2}{c|}{\dataDL}                                  & \multicolumn{2}{c|}{\dataDHL}                                 \\ \hline
\multicolumn{1}{|c|}{Param}                 & \multicolumn{1}{c|}{best-fit} & mean$\pm\sigma$               & \multicolumn{1}{c|}{best-fit} & mean$\pm\sigma$               & \multicolumn{1}{c|}{best-fit} & mean$\pm\sigma$               & \multicolumn{1}{c|}{best-fit} & mean$\pm\sigma$               \\ \hline
\multicolumn{1}{|c|}{$100~\omega{}_{b }$}   & \multicolumn{1}{c|}{2.246}    & $2.245_{-0.014}^{+0.014}$     & \multicolumn{1}{c|}{2.263}    & $2.262_{-0.013}^{+0.013}$     & \multicolumn{1}{c|}{2.243}    & $2.238_{-0.013}^{+0.013}$     & \multicolumn{1}{c|}{2.255}    & $2.256_{-0.014}^{+0.013}$     \\
\multicolumn{1}{|c|}{$100\,\theta{}_{s }$}   & \multicolumn{1}{c|}{1.042}    & $1.042_{-0.00026}^{+0.00029}$ & \multicolumn{1}{c|}{1.042}    & $1.042_{-0.00030}^{+0.00030}$ & \multicolumn{1}{c|}{1.042}    & $1.042_{-0.00030}^{+0.00029}$ & \multicolumn{1}{c|}{1.042}    & $1.042_{-0.00028}^{+0.00031}$ \\
\multicolumn{1}{|c|}{$ln10^{10}A_{s }$}     & \multicolumn{1}{c|}{3.046}    & $3.045_{-0.015}^{+0.013}$     & \multicolumn{1}{c|}{3.055}    & $3.056_{-0.017}^{+0.014}$     & \multicolumn{1}{c|}{3.046}    & $3.041_{-0.014}^{+0.013}$     & \multicolumn{1}{c|}{3.051}    & $3.052_{-0.013}^{+0.013}$     \\
\multicolumn{1}{|c|}{$n_{s }$}              & \multicolumn{1}{c|}{0.968}    & $0.9669_{-0.0039}^{+0.0035}$  & \multicolumn{1}{c|}{0.972}    & $0.9712_{-0.0037}^{+0.0036}$  & \multicolumn{1}{c|}{0.958}    & $0.9578_{-0.0033}^{+0.0036}$  & \multicolumn{1}{c|}{0.961}    & $0.9614_{-0.0035}^{+0.0032}$  \\
\multicolumn{1}{|c|}{$\tau{}_{\rm reio }$}  & \multicolumn{1}{c|}{0.057}    & $0.0564_{-0.0075}^{+0.0067}$  & \multicolumn{1}{c|}{0.063}    & $0.0627_{-0.0089}^{+0.0070}$  & \multicolumn{1}{c|}{0.054}    & $0.0531_{-0.0070}^{+0.0071}$  & \multicolumn{1}{c|}{0.059}    & $0.0584_{-0.0065}^{+0.0067}$  \\
\multicolumn{1}{|c|}{$\omega{}_{\rm cdm }$} & \multicolumn{1}{c|}{0.119}    & $0.1186_{-0.0009}^{+0.0009}$  & \multicolumn{1}{c|}{0.117}    & $0.1172_{-0.0009}^{+0.0008}$  & \multicolumn{1}{c|}{0.120}    & $0.1194_{-0.0008}^{+0.0009}$  & \multicolumn{1}{c|}{0.118}    & $0.1179_{-0.0009}^{+0.0007}$  \\
\multicolumn{1}{|c|}{$\Omega{}_{m }$}       & \multicolumn{1}{c|}{0.308}    & $0.307_{-0.005}^{+0.005}$     & \multicolumn{1}{c|}{0.308}    & $0.2978_{-0.0050}^{+0.0047}$  & \multicolumn{1}{c|}{0.313}    & $0.3123_{-0.0051}^{+0.0050}$  & \multicolumn{1}{c|}{0.303}    & $0.3023_{-0.0052}^{+0.0042}$  \\
\multicolumn{1}{|c|}{\Ho}                   & \multicolumn{1}{c|}{67.87}    & $67.94_{-0.40}^{+0.39}$       & \multicolumn{1}{c|}{68.70}    & $68.68_{-0.38}^{+0.40}$       & \multicolumn{1}{c|}{67.52}    & $67.55_{-0.38}^{+0.37}$       & \multicolumn{1}{c|}{68.26}    & $68.33_{-0.33}^{+0.40}$       \\
\multicolumn{1}{|c|}{$\sigma_8$}            & \multicolumn{1}{c|}{0.808}    & $0.8068_{-0.0057}^{+0.0051}$  & \multicolumn{1}{c|}{0.807}    & $0.8070_{-0.0062}^{+0.0056}$  & \multicolumn{1}{c|}{0.808}    & $0.8055_{-0.0058}^{+0.0049}$  & \multicolumn{1}{c|}{0.806}    & $0.8051_{-0.0056}^{+0.0052}$  \\
\multicolumn{1}{|c|}{$n_L$}                 & \multicolumn{1}{c|}{-2.303}   & $-2.304_{-0.0035}^{+0.0029}$  & \multicolumn{1}{c|}{-2.301}   & $-2.302_{-0.0034}^{+0.0029}$  & \multicolumn{1}{c|}{-2.311}   & $-2.312_{-0.0027}^{+0.0032}$  & \multicolumn{1}{c|}{-2.311}   & $-2.311_{-0.0027}^{+0.0030}$  \\
\multicolumn{1}{|c|}{$\Delta_L^2$}          & \multicolumn{1}{c|}{0.345}    & $0.3428_{-0.0054}^{+0.0055}$  & \multicolumn{1}{c|}{0.342}    & $0.3413_{-0.0059}^{+0.0054}$  & \multicolumn{1}{c|}{0.341}    & $0.3383_{-0.0054}^{+0.0053}$  & \multicolumn{1}{c|}{0.336}    & $0.3357_{-0.0052}^{+0.0055}$  \\ \hline
\end{tabular}
\caption{Parameter values from \lcdm fits.}
\label{tb:5}
\end{table*}

\begin{table*}[h!]
\centering
\begin{tabular}{c|cc|cc|cc|cc|}
\cline{2-9}
                                           & \multicolumn{2}{c|}{\dataD}                                  & \multicolumn{2}{c|}{\dataDH}                                  & \multicolumn{2}{c|}{\dataDL}                                  & \multicolumn{2}{c|}{\dataDHL}                                 \\ \hline
\multicolumn{1}{|c|}{Param}                & \multicolumn{1}{c|}{best-fit} & mean$\pm\sigma$              & \multicolumn{1}{c|}{best-fit} & mean$\pm\sigma$               & \multicolumn{1}{c|}{best-fit} & mean$\pm\sigma$               & \multicolumn{1}{c|}{best-fit} & mean$\pm\sigma$               \\ \hline
\multicolumn{1}{|c|}{$100~\omega{}_{b }$}  & \multicolumn{1}{c|}{2.252}    & $2.256_{-0.017}^{+0.017}$    & \multicolumn{1}{c|}{2.291}    & $2.288_{-0.016}^{+0.014}$     & \multicolumn{1}{c|}{2.239}    & $2.240_{-0.019}^{+0.014}$     & \multicolumn{1}{c|}{2.283}    & $2.280_{-0.013}^{+0.015}$     \\
\multicolumn{1}{|c|}{$100\,\theta{}_{s }$} & \multicolumn{1}{c|}{1.0420}   & $1.042_{-0.0003}^{+0.0003}$  & \multicolumn{1}{c|}{1.0424}   & $1.043_{-0.00028}^{+0.00033}$ & \multicolumn{1}{c|}{1.0420}   & $1.042_{-0.00031}^{+0.00041}$ & \multicolumn{1}{c|}{1.0427}   & $1.043_{-0.0003}^{+0.00028}$  \\
\multicolumn{1}{|c|}{$ln10^{10}A_{s }$}    & \multicolumn{1}{c|}{3.044}    & $3.046_{-0.015}^{+0.016}$    & \multicolumn{1}{c|}{3.042}    & $3.045_{-0.016}^{+0.015}$     & \multicolumn{1}{c|}{3.055}    & $3.053_{-0.015}^{+0.013}$     & \multicolumn{1}{c|}{3.054}    & $3.054_{-0.017}^{+0.013}$     \\
\multicolumn{1}{|c|}{$n_{s }$}             & \multicolumn{1}{c|}{0.9681}   & $0.9686_{-0.0040}^{+0.0038}$ & \multicolumn{1}{c|}{0.9713}   & $0.9714_{-0.0038}^{+0.0038}$  & \multicolumn{1}{c|}{0.9672}   & $0.9663_{-0.0034}^{+0.0042}$  & \multicolumn{1}{c|}{0.9718}   & $0.9711_{-0.0036}^{+0.0035}$  \\
\multicolumn{1}{|c|}{$\tau{}_{\rm reio }$} & \multicolumn{1}{c|}{0.0559}   & $0.0574_{-0.0078}^{+0.0079}$ & \multicolumn{1}{c|}{0.0599}   & $0.0610_{-0.0084}^{+0.0072}$  & \multicolumn{1}{c|}{0.0578}   & $0.0583_{-0.0079}^{+0.0062}$  & \multicolumn{1}{c|}{0.0630}   & $0.0638_{-0.0085}^{+0.0073}$  \\
\multicolumn{1}{|c|}{$\Delta N_\mathrm{eff}$}           & \multicolumn{1}{c|}{0.071}   & $0.24_{-0.17}^{+0.04}$      & \multicolumn{1}{c|}{0.55}   & $0.61_{-0.13}^{+0.13}$      & \multicolumn{1}{c|}{0.010}   & $0.19_{-0.18}^{+0.06}$      & \multicolumn{1}{c|}{0.68}   & $0.70_{-0.15}^{+0.13}$      \\
\multicolumn{1}{|c|}{$R_{\Gamma}$}         & \multicolumn{1}{c|}{0.0032}   & $0.0226_{-0.0226}^{+0.0054}$ & \multicolumn{1}{c|}{0.0082}   & $0.03_{-0.03}^{+0.01}$        & \multicolumn{1}{c|}{0.052}    & $0.05_{-0.01}^{+0.01}$        & \multicolumn{1}{c|}{0.055}    & $0.056_{-0.011}^{+0.009}$     \\
\multicolumn{1}{|c|}{$\omega_{\rm idm}$}   & \multicolumn{1}{c|}{0.1198}   & $0.123_{-0.003}^{+0.0016}$   & \multicolumn{1}{c|}{0.1276}   & $0.129_{-0.003}^{+0.0028}$    & \multicolumn{1}{c|}{0.1196}   & $0.124_{-0.004}^{+0.002}$     & \multicolumn{1}{c|}{0.1316}   & $0.1319_{-0.0028}^{+0.0029}$  \\
\multicolumn{1}{|c|}{$\Omega{}_{m }$}      & \multicolumn{1}{c|}{0.3061}   & $0.3059_{-0.0071}^{+0.0062}$ & \multicolumn{1}{c|}{0.2941}   & $0.2965_{-0.0058}^{+0.0056}$  & \multicolumn{1}{c|}{0.3124}   & $0.3134_{-0.0057}^{+0.0054}$  & \multicolumn{1}{c|}{0.3006}   & $0.3011_{-0.0054}^{+0.0052}$  \\
\multicolumn{1}{|c|}{\Ho}                  & \multicolumn{1}{c|}{68.38}    & $69.25_{-1.10}^{+0.78}$      & \multicolumn{1}{c|}{71.71}    & $71.80_{-0.85}^{+0.78}$       & \multicolumn{1}{c|}{67.60}    & $68.43_{-1.30}^{+0.66}$       & \multicolumn{1}{c|}{71.84}    & $71.86_{-0.90}^{+0.81}$       \\
\multicolumn{1}{|c|}{$\sigma_8$}           & \multicolumn{1}{c|}{0.8066}   & $0.8016_{-0.0082}^{+0.0110}$ & \multicolumn{1}{c|}{0.8144}   & $0.807_{-0.008}^{+0.010}$     & \multicolumn{1}{c|}{0.7864}   & $0.7885_{-0.0070}^{+0.0072}$  & \multicolumn{1}{c|}{0.7969}   & $0.7969_{-0.0071}^{+0.0075}$  \\
\multicolumn{1}{|c|}{$n_{L}$}              & \multicolumn{1}{c|}{-2.305}   & $-2.316_{-0.006}^{+0.013}$   & \multicolumn{1}{c|}{-2.306}   & $-2.317_{-0.008}^{+0.013}$    & \multicolumn{1}{c|}{-2.339}   & $-2.3360_{-0.0063}^{+0.0049}$ & \multicolumn{1}{c|}{-2.334}   & $-2.3350_{-0.0048}^{+0.0047}$ \\
\multicolumn{1}{|c|}{$\Delta_L^2$}         & \multicolumn{1}{c|}{0.342}    & $0.3312_{-0.0092}^{+0.0150}$ & \multicolumn{1}{c|}{0.345}    & $0.334_{-0.011}^{+0.014}$     & \multicolumn{1}{c|}{0.304}    & $0.3104_{-0.0087}^{+0.0076}$  & \multicolumn{1}{c|}{0.317}    & $0.3162_{-0.0076}^{+0.0070}$  \\ \hline
\end{tabular}
\caption{Parameter values from SIDR+ fits. Note that $R_{\Gamma}\sim 0.05$ is equivalent to $10^{7}\Gamma_0\sim 1.2$ in the alternative formulation of the interaction strength parameter.}
\label{tb:6}
\end{table*}


\begin{table*}[h!]
\centering
\begin{tabular}{c|cc|cc|cc|cc|}
\cline{2-9}
                                            & \multicolumn{2}{c|}{\dataD}                                   & \multicolumn{2}{c|}{\dataDH}                                  & \multicolumn{2}{c|}{\dataDL}                                  & \multicolumn{2}{c|}{\dataDHL}                                 \\ \hline
\multicolumn{1}{|c|}{Param}                 & \multicolumn{1}{c|}{best-fit} & mean$\pm\sigma$               & \multicolumn{1}{c|}{best-fit} & mean$\pm\sigma$               & \multicolumn{1}{c|}{best-fit} & mean$\pm\sigma$               & \multicolumn{1}{c|}{best-fit} & mean$\pm\sigma$               \\ \hline
\multicolumn{1}{|c|}{$100~\omega{}_{b }$}   & \multicolumn{1}{c|}{2.250}    & $2.256_{-0.016}^{+0.016}$     & \multicolumn{1}{c|}{2.283}    & $2.283_{-0.016}^{+0.017}$     & \multicolumn{1}{c|}{2.249}    & $2.255_{-0.017}^{+0.015}$     & \multicolumn{1}{c|}{2.281}    & $2.277_{-0.016}^{+0.015}$     \\
\multicolumn{1}{|c|}{$100*\,\theta{}_{s }$} & \multicolumn{1}{c|}{1.0422}   & $1.042_{-0.00046}^{+0.00036}$ & \multicolumn{1}{c|}{1.0433}   & $1.043_{-0.00037}^{+0.00039}$ & \multicolumn{1}{c|}{1.0422}   & $1.042_{-0.00039}^{+0.00039}$ & \multicolumn{1}{c|}{1.0432}   & $1.043_{-0.00036}^{+0.00037}$ \\
\multicolumn{1}{|c|}{$ln10^{10}A_{s }$}     & \multicolumn{1}{c|}{3.049}    & $3.051_{-0.015}^{+0.015}$     & \multicolumn{1}{c|}{3.057}    & $3.059_{-0.016}^{+0.016}$     & \multicolumn{1}{c|}{3.048}    & $3.045_{-0.014}^{+0.014}$     & \multicolumn{1}{c|}{3.050}    & $3.049_{-0.014}^{+0.015}$     \\
\multicolumn{1}{|c|}{$n_{s }$}              & \multicolumn{1}{c|}{0.9699}   & $0.9717_{-0.0060}^{+0.0048}$  & \multicolumn{1}{c|}{0.9838}   & $0.9835_{-0.0050}^{+0.0049}$  & \multicolumn{1}{c|}{0.9700}   & $0.9717_{-0.0052}^{+0.0047}$  & \multicolumn{1}{c|}{0.9829}   & $0.9806_{-0.0050}^{+0.0052}$  \\
\multicolumn{1}{|c|}{$\tau{}_{\rm reio }$}  & \multicolumn{1}{c|}{0.0567}   & $0.0574_{-0.0075}^{+0.0078}$  & \multicolumn{1}{c|}{0.0601}   & $0.0610_{-0.0083}^{+0.0074}$  & \multicolumn{1}{c|}{0.0567}   & $0.0557_{-00.0076}^{+0.0069}$ & \multicolumn{1}{c|}{0.0574}   & $0.0575_{-0.0075}^{+0.0068}$  \\
\multicolumn{1}{|c|}{$N_{\rm IR}$}          & \multicolumn{1}{c|}{0.12}     & $0.19_{-0.18}^{+0.05}$        & \multicolumn{1}{c|}{0.64}     & $0.63_{-0.17}^{+0.13}$        & \multicolumn{1}{c|}{0.12}     & $0.21_{-0.19}^{+0.05}$        & \multicolumn{1}{c|}{0.64}     & $0.59_{-0.15}^{+0.17}$        \\
\multicolumn{1}{|c|}{$\mathrm{log}\,z_t$}   & \multicolumn{1}{c|}{-4.20}    & $-4.46_{-0.19}^{+0.29}$       & \multicolumn{1}{c|}{-4.29}    & $-4.29_{-0.12}^{+0.09}$       & \multicolumn{1}{c|}{-4.20}    & $-4.33_{-0.07}^{+0.33}$       & \multicolumn{1}{c|}{-4.26}    & $-4.25_{-0.12}^{+0.15}$       \\
\multicolumn{1}{|c|}{$R_{\Gamma}$}          & \multicolumn{1}{c|}{0.067}    & $0.22_{-0.22}^{+0.04}$        & \multicolumn{1}{c|}{0.11}     & $0.15_{-0.11}^{+0.07}$        & \multicolumn{1}{c|}{0.052}    & $0.054_{-0.011}^{+0.010}$     & \multicolumn{1}{c|}{0.070}    & $0.07_{-0.01}^{+0.01}$        \\
\multicolumn{1}{|c|}{$\omega{}_{\rm idm}$}  & \multicolumn{1}{c|}{0.121}    & $0.123_{-0.003}^{+0.002}$     & \multicolumn{1}{c|}{0.129}    & $0.129_{-0.003}^{+0.003}$     & \multicolumn{1}{c|}{0.121}    & $0.122_{-0.003}^{+0.002}$     & \multicolumn{1}{c|}{0.129}    & $0.128_{-0.003}^{+0.003}$     \\
\multicolumn{1}{|c|}{$\Omega{}_{m }$}       & \multicolumn{1}{c|}{0.308}    & $0.308_{-0.007}^{+0.006}$     & \multicolumn{1}{c|}{0.295}    & $0.297_{-0.006}^{+0.006}$     & \multicolumn{1}{c|}{0.307}    & $0.305_{-0.006}^{+0.006}$     & \multicolumn{1}{c|}{0.294}    & $0.295_{-0.005}^{+0.005}$     \\
\multicolumn{1}{|c|}{\Ho}                   & \multicolumn{1}{c|}{68.42}    & $68.84_{-1.10}^{+0.71}$       & \multicolumn{1}{c|}{71.91}    & $71.75_{-0.91}^{+0.77}$       & \multicolumn{1}{c|}{68.55}    & $69.11_{-1.20}^{+0.69}$       & \multicolumn{1}{c|}{71.97}    & $71.63_{-0.89}^{+0.97}$       \\
\multicolumn{1}{|c|}{$\sigma_8$}            & \multicolumn{1}{c|}{0.7997}   & $0.793_{-0.012}^{+0.015}$     & \multicolumn{1}{c|}{0.8103}   & $0.805_{-0.013}^{+0.014}$     & \multicolumn{1}{c|}{0.8016}   & $0.804_{-0.007}^{+0.007}$     & \multicolumn{1}{c|}{0.8143}   & $0.812_{-0.007}^{+0.007}$     \\
\multicolumn{1}{|c|}{$n_{L}$}               & \multicolumn{1}{c|}{-2.348}   & $-2.448_{-0.037}^{+0.120}$    & \multicolumn{1}{c|}{-2.368}   & $-2.393_{-0.044}^{+0.077}$    & \multicolumn{1}{c|}{-2.338}   & $-2.338_{-0.006}^{+0.006}$    & \multicolumn{1}{c|}{-2.338}   & $-2.338_{-0.006}^{+0.006}$    \\
\multicolumn{1}{|c|}{$\Delta_L^2$}          & \multicolumn{1}{c|}{0.313}    & $0.273_{-0.035}^{+0.050}$     & \multicolumn{1}{c|}{0.310}    & $0.296_{-0.035}^{+0.041}$     & \multicolumn{1}{c|}{0.320}    & $0.322_{-0.008}^{+0.007}$     & \multicolumn{1}{c|}{0.328}    & $0.327_{-0.008}^{+0.007}$     \\ \hline
\end{tabular}
\caption{Parameter values from WZDR+ fits.}
\label{tb:7}
\end{table*}


\begin{table*}[]
\centering
\begin{tabular}{ccccccccccc}
\hline\hline
Datasets &
  Model &
  $\chi^2_{\rm Tot.}$ &
  $\chi^2_{\rm CMB}$ &
  $\chi^2_{\rm Pantheon}$ &
  $\chi^2_{\rm BAO}$ &
  $\chi^2_{\rm Pl. Lensing}$ &
  $\chi^2_{S_8}$ &
  $\chi^2_{\rm SH0ES}$ &
  $\chi^2_{\mathrm{full-shape}}$ &
  $\chi^2_{\mathrm{Ly}\alpha}$ \\ \hline
         & \lcdm & 4075.9 & 2769.4 & 1026.6 & 1.4 & 9.2  & 2.5 & -    & 266.8 & -    \\
         & SIDR  & 4073.3 & 2767.7 & 1025.7 & 1.5 & 9.3  & 2.2 & -    & 266.8 & -    \\
\dataD   & WZDR  & 4072.7 & 2766.7 & 1025.7 & 1.7 & 9.5  & 2.2 & -    & 266.9 & -    \\
         & SIDR+ & 4073.4 & 2768.2 & 1025.9 & 1.5 & 9.4  & 1.9 & -    & 266.5 & -    \\
         & WZDR+ & 4071.1 & 2767.2 & 1025.9 & 1.4 & 9.2  & 1.3 & -    & 266.0 & -    \\ \hline
         & \lcdm & 4102.8 & 2772.7 & 1025.6 & 2.3 & 9.9  & 0.5 & 26.1 & 265.6 & -    \\
         & SIDR  & 4082.4 & 2773.0 & 1025.7 & 2.8 & 10.3 & 0.8 & 2.8  & 267.0 & -    \\
\dataDH  & WZDR  & 4080.4 & 2768.7 & 1025.8 & 3.0 & 10.4 & 1.8 & 2.2  & 268.4 & -    \\
         & SIDR+ & 4082.6 & 2773.7 & 1025.7 & 2.6 & 10.1 & 0.8 & 2.9  & 266.7 & -    \\
         & WZDR+ & 4078.8 & 2770.4 & 1025.7 & 2.6 & 10.0 & 0.6 & 2.1  & 267.3 & -    \\ \hline
         & \lcdm & 4103.7 & 2774.5 & 1026.2 & 1.1 & 8.6  & 3.8 & -    & 267.4 & 22.1 \\
         & SIDR  & 4101.4 & 2774.2 & 1026.0 & 1.1 & 8.9  & 3.0 & -    & 267.5 & 20.7 \\
\dataDL  & WZDR  & 4101.1 & 2775.0 & 1025.9 & 1.3 & 9.0  & 2.7 & -    & 266.8 & 20.3 \\
         & SIDR+ & 4072.2 & 2769.3 & 1026.0 & 1.1 & 9.3  & 0.5 & -    & 265.9 & 0.1  \\
         & WZDR+ & 4071.7 & 2767.6 & 1025.8 & 1.5 & 9.5  & 1.3 & -    & 265.9 & 0.3  \\ \hline
         & \lcdm & 4138.2 & 2779.4 & 1025.7 & 1.7 & 9.3  & 1.2 & 31.2 & 265.6 & 24.0 \\
         & SIDR  & 4111.7 & 2781.5 & 1025.8 & 2.3 & 9.5  & 1.6 & 2.7  & 266.6 & 21.6 \\
\dataDHL & WZDR  & 4115.7 & 2781.1 & 1025.7 & 2.0 & 9.2  & 2.1 & 9.0  & 266.8 & 19.8 \\
         & SIDR+ & 4087.9 & 2780.1 & 1025.6 & 2.0 & 10.5 & 0.2 & 2.1  & 266.6 & 0.9  \\
         & WZDR+ & 4080.0 & 2769.9 & 1025.7 & 2.8 & 10.5 & 0.8 & 2.0  & 267.5 & 0.9  \\ \hline\hline
\end{tabular}
\caption{Minimum $\chi^2$ for all datasets and cosmological models. \lya refers to the eBOSS Lyman-alpha forest likelihood.}
\label{tb:8}
\end{table*}

\end{document}